\documentclass{article}
\usepackage{graphicx}
\usepackage{xcolor}
\usepackage{amsfonts}
\usepackage{amssymb}
\def\ligne#1{\hbox to\hsize{#1}}
\def\leurre{\noindent\leftskip0pt\small\baselineskip 10pt}
\newtheorem{thm}{\textbf{Theorem}}
\newtheorem{fig}{\textbf{Figure}}
\newtheorem{tab}{\textbf{Table}}

\author{Maurice {\sc Margenstern}}
\title{An outer totalistic weakly universal cellular automaton in the dodecagrid with 
four states}
\begin{document}
\maketitle

\begin{abstract}
In this paper, we prove that there is an outer totalistic weakly universal cellular 
automaton in the dodecagrid, the tessellation $\{5,3,4\}$ of the hyperbolic $3D$ space, 
with four states. It is the first result in such a context. 
\end{abstract}

\section{Introduction}~\label{intro}

    In many papers, the author studied the possibility to construct universal cellular
automata in tilings of the hyperbolic plane, a few ones in the hyperbolic $3D$ space. 
Most often, the constructed cellular automaton was weakly universal. By {\it weakly 
universal}, we mean that the automaton is able to simulate a universal device starting
from an infinite initial configuration. However, the initial configuration should not be
arbitrary. It was the case that it was periodic outside a large enough circle, in fact
it was periodic outside such a circle in two different directions as far as the simulated
device was a two-registered machine. From a result by Minsky, \cite{minsky}, it is enough to simulate any Turing machine. In almost all papers, the considered tiling of the
hyperbolic plane was either the pentagrid or the heptagrid, {\it i.e.} the tessellation
$\{5,4\}$, $\{7,3\}$ respectively.  Both tessellations live in the hyperbolic plane only.
In the pentagrid, the basic tile is a regular convex pentagon with right angles. In the
heptagrid, it is a regular convex heptagon with the angle 
\hbox{$\displaystyle{{2\pi}\over3}$} between consecutive sides. 
In the present paper, we consider the dodecagrid which we define and explain in 
Sub-section~\ref{ssdodec}. In Sub-section~\ref{sstot}, we remind the read what 
outer totalistic cellular automata are.

\subsection{The dodecagrid}\label{ssdodec}

In the present paper, we consider the tiling of the hyperbolic $3D$-space which we 
call the {\bf dodecagrid} whose signature is defined as \hbox{$\{5,3,4\}$}. 
In that sig\-nature, \goodbreak\noindent 5 is the number of sides of a face, 3 is the 
number of edges which meet at a vertex, 4 is the number of dodecahedrons around an edge. 
In the hyperbolic 
$3D$-space, there is another tessellation based on another dodecahedron whose signature is
\hbox{$\{5,3,5\}$} which means that around an edge, there should be five dodecahedrons.
From now on, we consider the dodecagrid only and its dodecahedrons are always copies of 
Poincar\'e's dodecahedron we denote by $\Delta$.

Below, Figure~\ref{fdodecs}
provides us with a representation of the dodecagrid according to Schlegel representation
of the solid. We shall more frequently use another representation illustrated by
Figure~\ref{stab_fix0} which we shall again meet later in Section~\ref{scenario}.
That figure makes use of a property we shall see with the explanations about 
Figure~\ref{fdodecs}. Before turning to that argumentation, we presently deal with the 
Schlegel representation of alone $\Delta$.

   Figure~\ref{fdodecs} is obtained by the projection of the vertices and the edges of
a dodecahedron of the hyperbolic $3D$-space on the plane of one of its faces. 
\vskip 10pt
\vtop{
\ligne{\hfill
\includegraphics[scale=0.4]{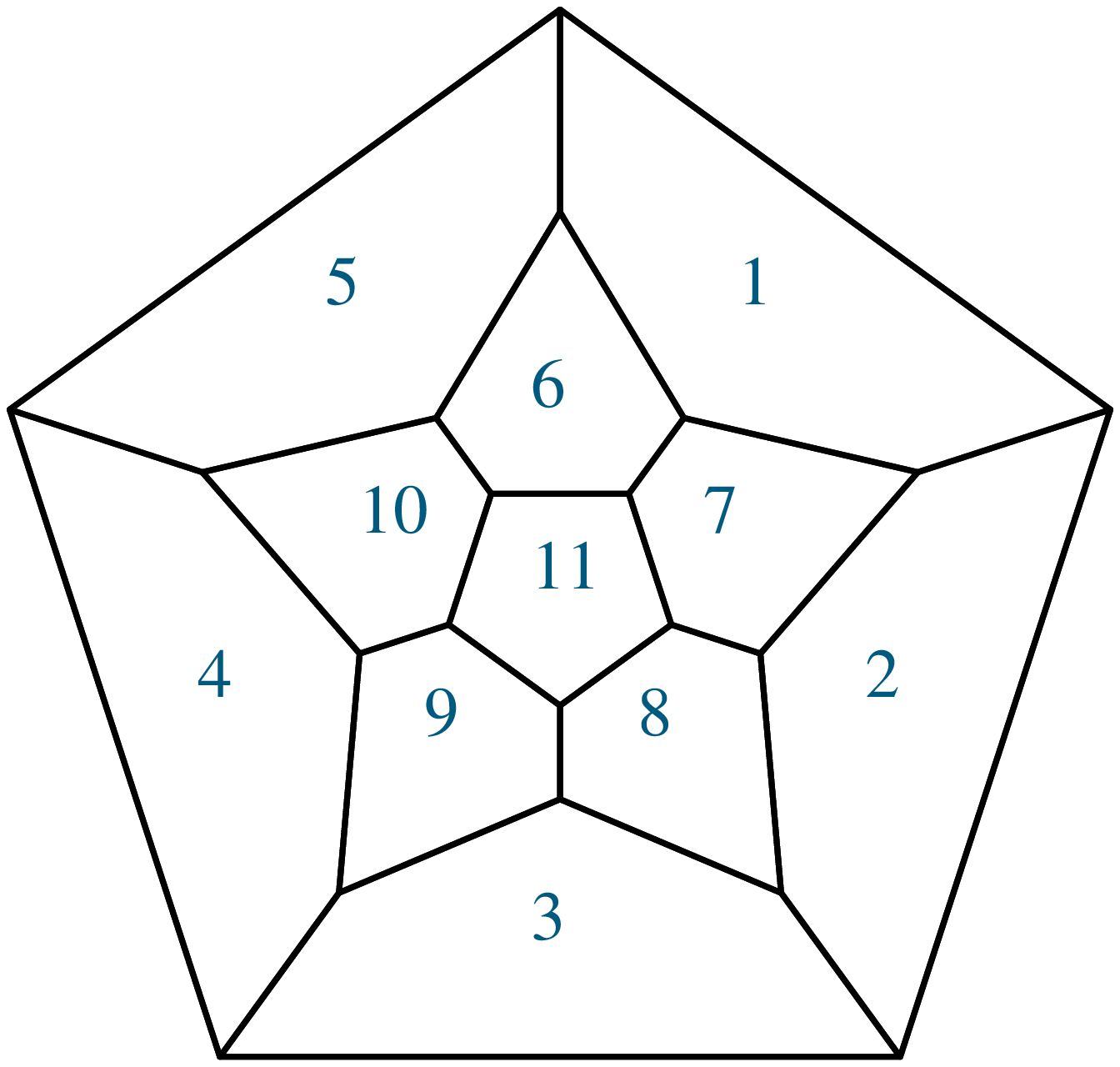}
\hfill}
\vspace{-5pt}
\begin{fig}\label{fdodecs}
\leurre
The Schlegel representation of the dodecahedron~$\Delta$.
\end{fig}
}

As shown on the figure, we number the faces of~$\Delta$. Face~0 is the face whose plane 
is the one on which the projection is performed. We number the faces around face~0 by 
clockwise turning around the face when we look at the plane from a point which stands 
over~$\Delta$. 'Over' means that $\Delta$ and the centre of the projection are on the 
same half-space defined by the plane on which the projection is performed. If we 
consider the face which is opposed to face~0, we also number the remaining faces 
of~$\Delta$, they are also clockwise numbered from~6 to~10 with face~6 sharing a side 
with face~5 and another one with face~1. The not yet numbered face which is opposite to 
face~0 is now numbered~11. That numbering will be the basic numbering from which we 
derive a numbering of the vertices and of the edges as follows: a vertex or an edge
will be denoted by the numbers of the faces which share that element, two numbers for
an edge, three ones for a vertex.

There is a problem with that representation of which the reader should be aware. 
Put a dodecahedron~$\Delta_i$ on each face $i$ of~$\Delta$. Assume that we number the 
faces of $\Delta_i$ by numbering with~0 the face of each dodecahedron $\Delta_i$ which 
lies on the face~$i$ of~$\Delta$ and denote those dodecahedrons by 
${\Delta_i}_0$ respectively. Consider $\Delta_i$ and $\Delta_j$ whose faces~0 are
contiguous faces of~$\Delta$, which means that both dodecahedrons share the edge shared
by the faces~$i$ and~$j$ of~$\Delta$ with, of course \hbox{$i\not= j$}. We may assume the 
numbering of the faces of those dodecahedrons is such that the faces~1 of~$\Delta_i$ 
and~$\Delta_j$ also share the side $i$-$j$. We are with three dodecahedrons around the 
edge $i$-$j$: $\Delta_i$, $\Delta_j$ and~$\Delta$. If we put a dodecahedron 
$\Delta_{ij}$, $\Delta_{ji}$ on the face~1 of~$\Delta_i$, $\Delta_j$ respectively, then, 
necessarily, we have \hbox{$\Delta_{ij}=\Delta_{ji}$} as far as there must be exactly 
four dodecahedrons around the side $i$-$j$.

\def\HH{$\mathcal H$}
   The cellular automaton we construct in Section~\ref{scenario} evolves in the
hyperbolic $3D$ space but a large part of the construction deals with a single plane
which we shall call the {\bf horizontal plane} denoted by \HH. In fact, both sides
of~\HH{} will be used by the construction and most tiles of our construction have a face
on~\HH.

   We take advantage of that circumstance to define another representation, possibly more 
convenient for our purpose. 

   The trace of the dodecagrid on~\HH{} is a tiling of the hyperbolic, namely the
tiling $\{5,4\}$ we call the {\bf pentagrid}. The left-hand side part of 
Figure~\ref{penta} illustrates the tiling and its right-hand side part illustrates a way 
to locate the cells of the pentagrid. Let us look closer at the figure whose pictures 
live in Poincar\'e's disc, a popular representation of the hyperbolic plane, 
see~\cite{mmbook1}.

   In the left-hand side picture, we can see five tiles which are counter-clockwise
numbered from~1 up to~5, those tiles being the neighbours of a tile which we call
the {\bf central tile} for convenience. Indeed, there is no 
central tile in the pentagrid
as there is no central point in the hyperbolic plane. We can see the disc model as a
window over the hyperbolic plane, as if we were flying over that plane in an abstract
spacecraft. The centre of the circle is the point on which are attention is focused while
the circle itself is our horizon. Accordingly, the central tile is the tile which is 
central with respect to the area under our consideration. It is also the reason to
number the central tile by~0.

The right-hand side picture shows us five blocs of tiles we call {\bf sectors}. 
Each sector is defined by a unique tile which shares and edge with the central one.
We number that tile by~1. It is a green tile on the picture. The sector is delimited
by two rays~$u$ and~$v$ issued from a vertex of tile~1: they continue two consecutive 
sides of tile~0.  Those rays are supported by straight lines in the hyperbolic plane 
and they define a right angle. Tile~1 is called the {\bf head} of the sector it defines :
the sector is the set of tiles contained in the angle defined by~$u$ and~$v$.
We also number the sectors from~1 to~5 by counter-clockwise turning around tile~0.
We also say that two tiles are {\bf neighbouring} or that they are {\bf neighbours} of 
each other if and only if they have a common side.

\vskip 10pt
\vtop{
\ligne{\hfill
\includegraphics[scale=0.75]{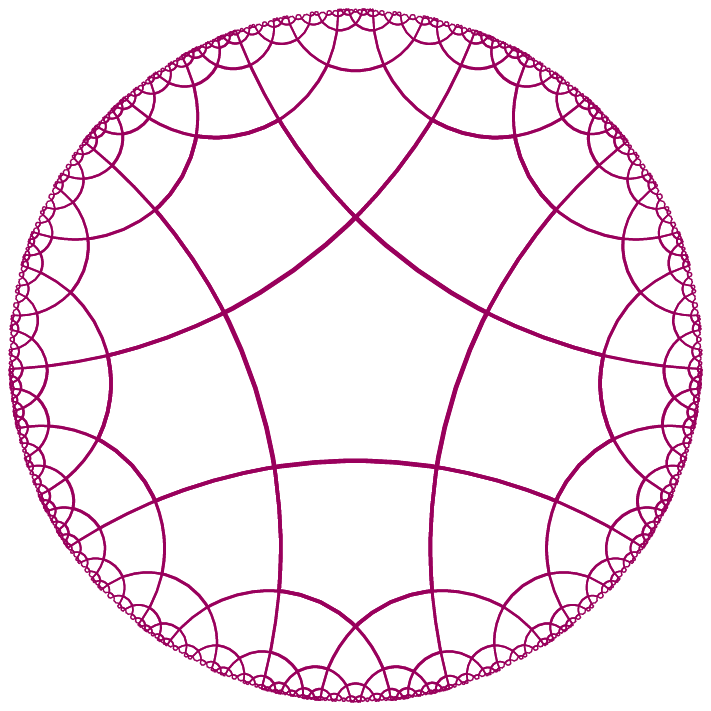}
\raise-20pt\hbox{\includegraphics[scale=0.475]{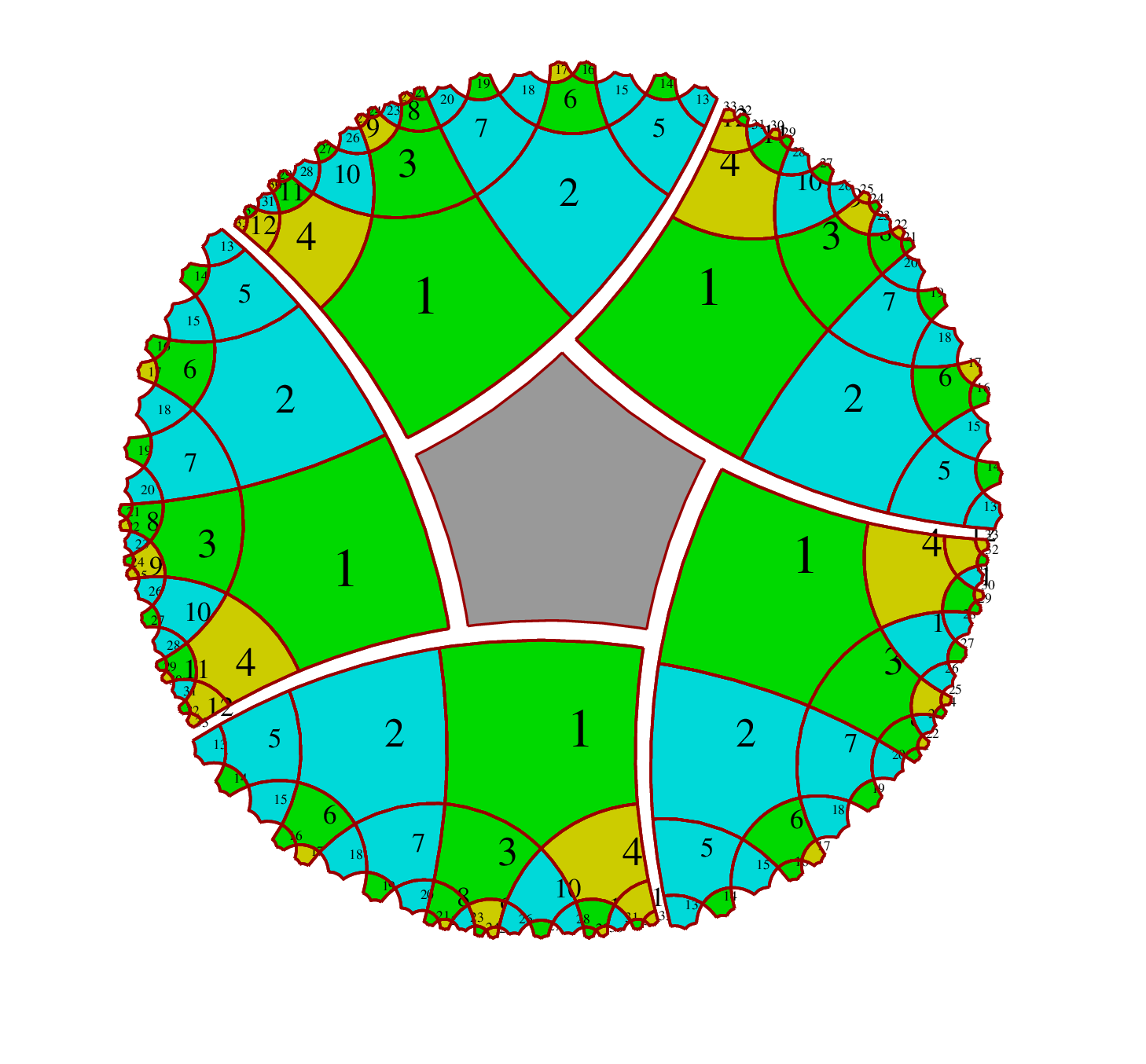}}
\hfill}
\vspace{-5pt}
\begin{fig}\label{penta}
\leurre
To left: a representation of the pentagrid in Poincar\'es disc model of the hyperbolic 
plane. To right: a splitting of the hyperbolic plane into five sectors around a once
and for all fixed central tile.
\end{fig}
}

   Consider the configuration illustrated by Figure~\ref{stab_fix0}. We can see the 
Schlegel projection of a dodecahedron on each tile of the picture. A few dodecahedrons
have another light colour and on each of them, some faces are green and a few ones are 
red. We call them the elements of a {\bf track}.  
\vskip 10pt
\vtop{
\ligne{\hfill
\includegraphics[scale=1.25]{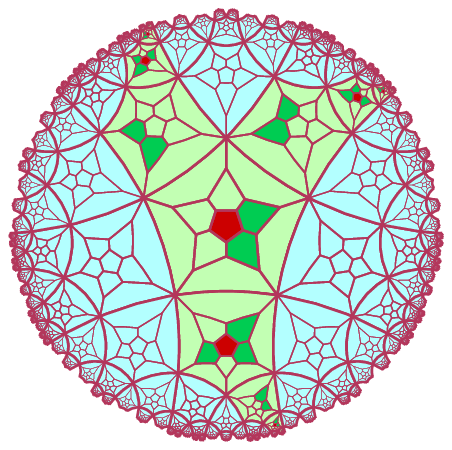}
\hfill}
\vspace{-5pt}
\begin{fig}\label{stab_fix0}
\leurre
A configuration we shall later meet in Section~{\rm\ref{scenario}}. The central tile
is, in some sense, the centre of that configuration. A green, red face means that a
green, red dodecahedron respectively is put on that face.
\end{fig}
}

Tile~0 is the central tile of the 
picture. Around tile~0, we can see three elements of track exactly, the other i
neighbouring tiles having a uniform light colour: we call them blank tiles. We number 
the elements of tracks by the number of their sectors, here 1, 3 and~5 by counter-clockwise
turning around tile~0. 
On the picture, tile~3 is just below tile~0. Let us look closer at tiles~0 and~3 of~\HH. Let $\Delta_0$ and~$\Delta_3$ be the tiles of the dodecagrid which 
are put on tiles~0 and~3 of~\HH{} respectively.  Assume that the face~1 of $\Delta_0$ and 
of~$\Delta_3$ share the side which is shared by the tiles~0 and~3 of~\HH. By construction
of the dodecagrid, below~\HH, there are also two dodecahedrons, one below~$\Delta_0$, 
the other below~$\Delta_3$. Accordingly, the face~1 of~$\Delta_0$ and that of~$\Delta_3$ 
coincide as far as four dodecahedrons share the common side of tiles~0 and~3 of~\HH.  
Using the numbering of the faces of~$\Delta_0$ and~$\Delta_3$ just above defined, we can 
see that if we put a dodecahedron on the faces~6 of~$\Delta_0$ and~$\Delta_3$ those 
dodecahedrons share a face which is on the same plane as the common face~1 of~$\Delta_0$ 
and~$\Delta_3$. Consider, for instance, the faces~6 and~7 of~$\Delta_0$ which also share a common side of~$\Delta_0$. Put a dodecahedron $D_i$ on the face~$i$ of $\Delta_0$. Then 
$D_6$ and~$D_7$ have no common face: their common side is shared by a fourth dodecahedron
outside $\Delta_0$, $D_6$ and $D_7$ as just indicated. Accordingly, we have to pay 
attention to dodecahedrons which are put on the faces of a dodecahedron in the 
representation as defined on Figure~\ref{stab_fix0}: we call that representation the 
{\bf \HH-representation}. The rule is simple: two dodecahedrons sharing an edge~$s$ in 
the \HH-representation also share their face sharing~$s$ if and only if $s$ is also a 
common side of the tiles of~\HH{} on which those faces are projected. 

As mentioned in the caption of Figure~\ref{stab_fix0}, a dark blue face of the 
projection of a dodecahedron means that, in the dodecagrid, a dark blue dodecahedron is 
put in that face. By {\it abus de langage}, we also call \HH{} the restriction of the 
dodecagrid to those which sit on that plane. When it will be needed to clarify, we denote 
by \HH$_u$, \HH$_b$ the set of dodecahedrons which are placed upon, below~\HH{} 
respectively.

\subsection{Outer totalistic cellular automata}\label{sstot}

   Let us remind the reader that cellular automata are a model of massive parallelism.
The base of a cellular automaton is a cell. The set of cells is supposed to be homogeneous
in several aspects: the neighbours of each cell constitute subsets which have the same
structure; the cell changes its state at each tip of a discrete clock according the
states of its neighbours and its own state. The change is dictated by a finite automaton 
which is the same for each cell. A regular tiling is an appropriate space for implementing
cellular automata: a cell is the combination of a tile together with the finite automaton
ruling the change of states. The tile is called the {\bf support} of the cell. Let $T$ 
be a tile and let $N(T)$ be the set of its neighbours. By regular, we mean that the 
number of elements of $N(T)$ is the same for any~$T$. The dodecagrid satisfies that 
requirement. Moreover, there is an algorithm to locate the tiles which is cubic in time 
in the size of the code attached to each tile, see \cite{mmbook2} for instance. However, 
taking benefit of~\HH, we inherit of the linear algorithm allowing us to locate the
tiles of~\HH. That use is reinforced by the face that our incursions in the third 
dimension will not lead us far from~\HH. From now on we indifferently say tile or cell 
for a dodecahedron of the dodecagrid, confusing the cell with its support.

    The way the automaton manages the change of states can be defined by a finite set
of {\bf rules} we shall call the {\bf program} of the automaton and we shall organise
it in a {\bf table} we shall display by pieces only. In Section~\ref{srules}, we define
the format of the rules. However we need not enter such details in order to define
outer totalistic cellular automata. The alphabet $\mathbb A$ of the automaton attached 
to each cell is the set of the possible states taken by the cell. Define an order on
$\mathbb A$ so that we can write \hbox{${\mathbb A} = \{e_0,...,e_n\}$}, so that $n$+1
is the number of states. By definition, the index of the state $e_i$ is~$i$.
The cellular automaton is said to be {\bf outer totalistic} if the state at time~$t$+1
of a cell~$\kappa$ exactly depends both on the state at time~$t$ of~$\kappa$ and on 
the sum of the indices of the states at time~$t$ of all $\lambda$ for 
$\lambda\in N(\kappa)$, we call it the {\bf weight} of the neighbours. The cellular 
automaton is said {\bf totalistic} if the state at time $t$+1 of~$\kappa$ depends on 
the sum of weight of the cell at time~$t$ and of the state of~$\kappa$ at time~$t$. The 
maximal value for the weight of the neighbours of a cell~$\kappa$ is 
\hbox{$n$+1$\times\vert N(\kappa)\vert$}. Call that sum the {\bf maximal weight} for a 
rule.

   Now that the global setting is given, we shall proceed as follows: 
Section~\ref{scenario} indicates the main lines of the implementation which is precisely
described in Subsection~\ref{newrailway}. At last, Section~\ref{srules} gives us the rules
followed by the automaton. That section also contain a few figures which illustrate the 
application of the rules. Those figures were established from pieces of figures drawn by 
a computer program which applied the rules of the automaton to an appropriate window in 
each of the configurations described in Subsection~\ref{newrailway}. The computer 
program also computed the sum of states of the neighbour of a cell.

   That allowed us to prove the following property:

\begin{thm}\label{letheo}
There is a weakly universal cellular automaton in the dodecagrid
which is outer totalistic and truly spatial. Its automaton has four states and the
maximal sum of states in the neighbourhood of a cell is $21$, less than the half
of the maximal weight for a rule.
\end{thm}

\section{Main lines of the computation}\label{scenario}

   The first paper about a universal cellular automaton in the pentagrid, the 
tessellation $\{5,4\}$ of the hyperbolic plane, was \cite{fhmmTCS}. That cellular 
automaton was also rotation invariant, at each step of the computation, the set of non 
quiescent states had infinitely many cycles: we shall say that it is a truly planar 
cellular automaton. That automaton had 22~states. That result was improved by a cellular 
automaton with 9~states in~\cite{mmysPPL}. Recently, it was improved with 5~states, 
see~\cite{mmpenta5st}. A bit later, I proved that in the heptagrid, the tessellation 
$\{7,3\}$ of the hyperbolic plane, there is a weakly universal cellular automaton with 
three states which is rotation invariant and which is truly planar, \cite{mmhepta3st}. 
Later, I improved the result down to two states but the rules are no more rotation 
invariant, see~\cite{mmpaper2st}. Paper \cite{JAC2010} constructs three cellular
automata which are strongly universal and rotation invariant: one in the pentagrid, one 
in the heptagrid, one in the tessellation \hbox{$\{5,3,4\}$} of the hyperbolic 
$3D$-space. By strongly universal we mean that the initial configuration is finite, 
{\it i.e.} it lies within a large enough circle.

    In the present paper, as we go back to weak universality, we take the general
frame of the quoted papers. For the convenience of the reader, we repeat the main ideas
in Sub-section~\ref{railway} and we focus on the specificity of the present paper
in Sub-subsection~\ref{newrailway}.

\subsection{The railway model}\label{railway}

    The simulation is based on the railway model devised in~\cite{stewart} which lives 
in the Euclidean plane. It consists of {\bf tracks} and {\bf switches} and the 
configuration of all switches at time~$t$ defines the configuration of the computation 
at that time. There are three kinds of switches, illustrated by Figure~\ref{switches}. 
The changes of the switch configurations are performed by a locomotive which runs over 
the circuit defined by the tracks and their connections organised by the switches.

A switch gathers three tracks $a$, $b$ and~$c$ at a point. In an active crossing,
the locomotive goes from~$a$ to either~$b$ or~$c$. In a passive crossing, it goes
either from~$b$ or~$c$ to~$a$. 

\vskip 10pt
\vtop{
\ligne{\hfill
\includegraphics[scale=0.8]{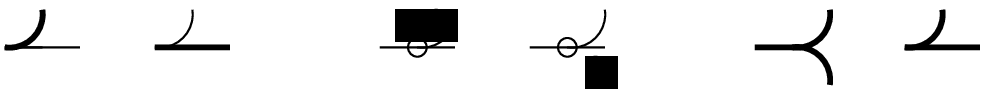}
\hfill}
\begin{fig}\label{switches}
\leurre
The switches used in the railway circuit of the model. To left, the fixed switch, in the
middle, the flip-flop switch, to right the memory switch. In the flip-flop switch, the 
bullet indicates which track has to be taken.
\end{fig}
}

In the fixed switch, the locomotive goes from~$a$ 
to always the same track: either~$b$ or~$c$. The passive crossing of the fixed switch is
possible. The flip-flop switch is always crossed actively only. If the locomotive
is sent from~$a$ to~$b$, $c$ by the switch, it will be sent to~$c$, $b$ respectively at 
the next passage. The memory switch can be crossed actively or passively. Now, the track 
taken by the locomotive in an active passage is the track taken by the locomotive in the 
last passive crossing.

\vtop{
\ligne{\hfill
\includegraphics[scale=0.6]{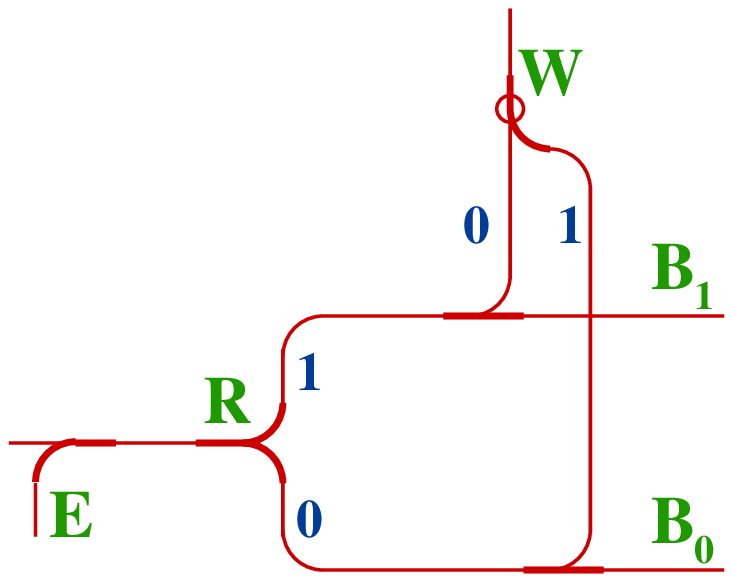}
\hfill}
\begin{fig}\label{basicelem}
\leurre
The basic element containing one bit of information.
\end{fig}
}

Figure~\ref{basicelem} illustrates the circuit which stores a one-bit unit of 
information. The locomotive may enter the circuit either through the gate~$R$ or through 
the gate~$W$.

  If it enters through the gate~$R$ where a memory switch sits, it goes either through
the track marked with~1 or through the track marked with~0. When it crossed the switch
through track~1, 0, it leaves the unit through the gate~$B_1$, $B_0$ respectively.
Note that on both ways, there are fixed switch sending the locomotive to the appropriate
gate~$B_i$. Note that when the locomotive leaves the unit, no switch was changed.
If the locomotive enters the unit through the gate~$W$, it is sent to the 
gate~$R$, either through track~0 or track~1 from~$W$. Accordingly, the locomotive
arrives to~$R$ where it crosses the switch passively, leaving the unit through the 
gate~$E$ thanks to a fixed switch leading to that latter gate. When the locomotive 
took track~0, 1 from~$W$, the switch after that indicates track~1, 0 respectively and the 
locomotive arrives at~$R$ through track~1, 0 of~$R$. The tracks are numbered according to 
the value stored in the unit. Note that when the locomotive leaves the unit, two switches
were changed: the flip-flop at~$W$ and the memory switch at~$R$. 

By definition, the unit is~0, 1 when both tracks from~$W$ and from~$R$ are~0, 1 
respectively. So that, as seen from that study, the entry through~$R$ performs a reading 
of the unit while the entry through~$W$, changes the unit from~0 to~1 or from~1 to~0: the 
entry through~$W$ should be used when it is needed to
change the content of the unit and only in that case. The structure works like a memory
which can be read or rewritten. It is the reason why call it the {\bf one-bit memory}.

   We shall see how to combine one-bit memories in the next sub-section as far as we 
introduce several changes to the original setting for the reasons we indicate there.

\subsection{Tuning the railway model}\label{newrailway}

   We first look at the implementation of the tracks in Sub-subsection~\ref{ssstracks}
and how it is possible to define the crossing of two tracks.
In Sub-subsection~\ref{sssswitch} we see how the switches are implemented.
Then, in Sub-subsection~\ref{sssunit}, we see how the one-bit memory is implemented in 
the new context and then, in Sub-section~\ref{sssregdisp}, how we use it in various 
places. At last but not the least, we shall indicate how registers are implemented
in Sub-subsection~\ref{sssreg}.

\subsubsection{The tracks}\label{ssstracks}

    The tracks play a key role in the computation, as important as instructions and 
registers: indeed, they convey information without which any computation is impossible.
Moreover, as can be seen in many papers of the author, that one included, it is not an
obvious issue which must always be addressed.
  
    It is not useful to list the similarities and the distinctions between the present 
implementation and those of my previous papers. The best is to focus on the implementation
used by this paper. If the reader is interested by the comparison with previous 
implementations the references already indicated give him/her access to the corresponding
papers.

\newcount\compterel\compterel=1
\def\numerrel{\the\compterel\global \advance\compterel by 1}
\def\ftt #1 {{\footnotesize\tt#1}}
\def\VV{\hbox{$\mathcal V$}}
   The tracks are one-way. It is already needed by the constraint of outer totalisticity.
Indeed, in previous papers, the elements of tracks consist of a blank cell with a 
few non-blank neighbours placed at appropriated places. We indicate those non-blank
neighbours as well as their places by the word {\bf decoration}. 
In previous papers, the motion was organised according to the following scheme:
\vskip 5pt
\ligne{\hfill\ftt{
   WWW\hskip 20pt LWW\hskip 20pt WLW\hskip 20pt WWL}
\hfill(\numerrel)\hskip 10pt}
\vskip 5pt
Under the constraint of outer totalisticity, such a scheme cannot work: if all elements
of the track are identical and if the moving state is always the same, the sum of
indices of the non-blank states in the neighbourhood is the same in all the above patterns
so that as far as \ftt{L } occurs in a cell, it is seen by two neighbouring cells.
Accordingly, \ftt{L } will appear in both neighbours of the cell, which is not what is
expected. Accordingly, we replace (1) by the following scheme :
\vskip 5pt
\ligne{\hfill\ftt{
   3B234234\hskip 20pt 34R34234\hskip 20pt 342G4234\hskip 20pt 3423B234\hskip 20pt 
3423B234\hskip 20pt 342342R4}
\hfill(\numerrel)\hskip 10pt}
\vskip 5pt
The motion is represented by a moving locomotive which successively takes different 
colours in consecutive cells which differ by their decorations according to a
periodic pattern. The successive pattern is defined by \ftt{234 } where 2, 3 and 4 refer
to the number of non-blank cells in the corresponding decorations of the cells
as illustrated by Figure~\ref{ftracks}.

Now, as soon will be seen, we need a two-way circulation in some portions of the circuit.
It is a point where the third direction comes to help us. In many portions, the
circuit can be implemented on a fixed plane we call \HH, already mentioned in the 
introduction. Roughly speaking, the traffic in one direction will occur 
in~\HH$_u$, \HH$_b$ while the reverse running will be performed in~\HH$_b$, \HH$_u$ 
respectively. Occasionally and locally, we shall use a plane \VV{} which is 
orthogonal to~\HH. 

The present organisation of the one-way tracks is very different from all the 
implementations described in the previous papers of the author. As already mentioned,
we mark the elements of a track by milestones whose set constitutes the decoration of the
element. But here, the decorations are different and their succession is performed 
according to the periodic repetition of the same pattern, namely \ftt{234 }, as already
indicated. As in~\cite{mmarXiv21b}, the return motion along a track in \HH$_u$, \HH$_b$
is performed along another track which is placed in \HH$_b$, \HH$_u$ respectively. But,
contrarily to~\cite{mmarXiv21b}, the elements of the return track does not share its
face~0 with that of the direct track. In order to make things more accurate, we say
that a {\bf track} is a sequence of elements such that its consecutive members 
can see each other but it cannot see other elements outside its neighbours in the 
sequence. Moreover, we require that each element of the track has a side on a 
line of~\HH{} which indicates the direction of the track. We say that the track 
{\bf follows} that line. We define a {\bf path} to be
a sequence of tracks which are linked in some way which will be later explained.
Elements in a path are imposed the same conditions as elements in a track with respect
to the elements they can see. The left-hand side part of Figure~\ref{fpaths} illustrates 
a track while its right-hand side part illustrates the same track together with its 
return track.

\vskip 10pt
\vtop{
\ligne{\hfill
\includegraphics[scale=0.4]{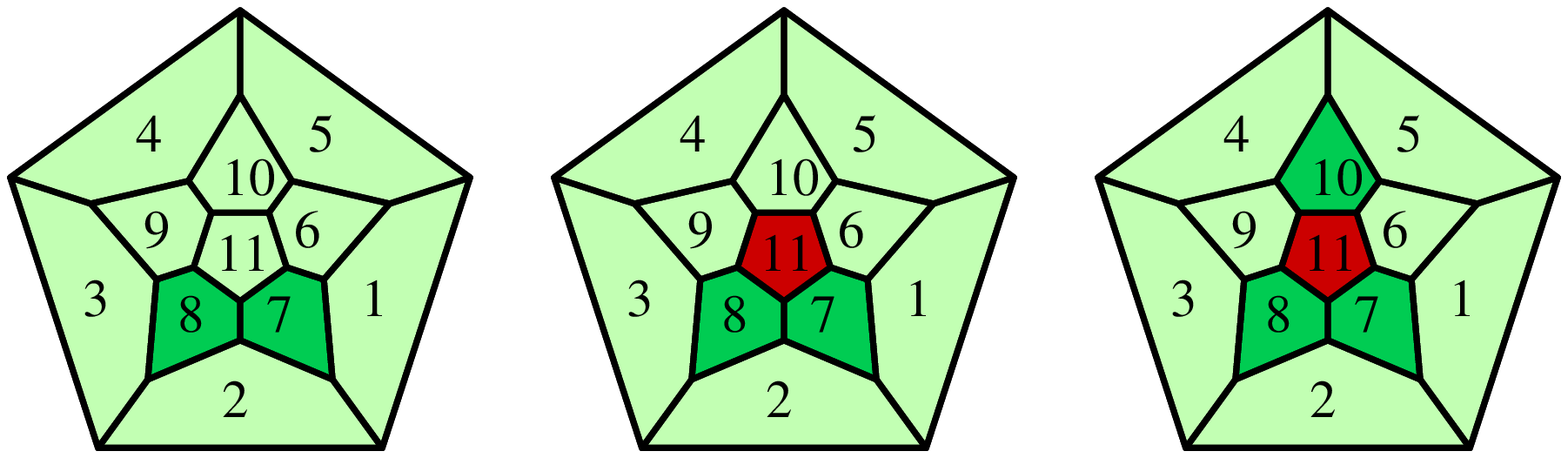}
\hfill}
\begin{fig}\label{ftracks}
\leurre
Three consecutive elements of a track illustrating the pattern \ftt{234 } .
\end{fig}
}

\vskip 10pt
\vtop{
\ligne{\hfill
\includegraphics[scale=0.8]{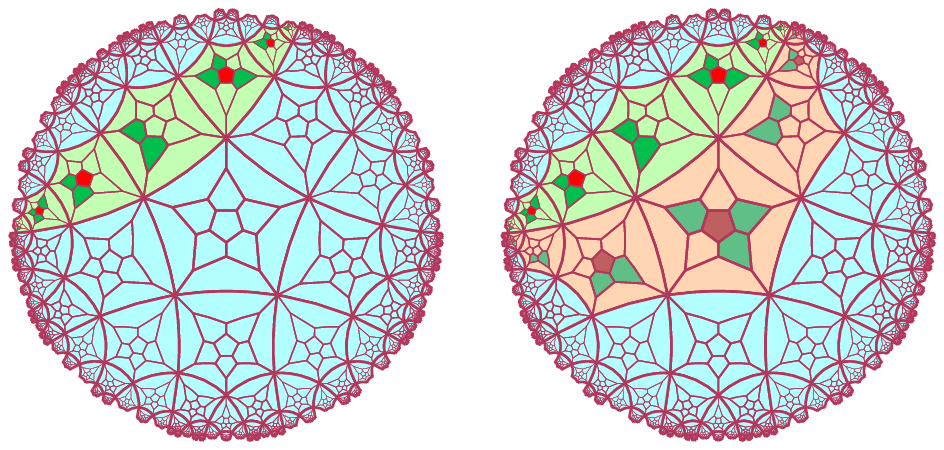}
\hfill}
\vskip-20pt
\begin{fig}\label{fpaths}
\leurre
To left, a track, to right its return track. They both follow the same line.
\end{fig}
}

On that latter part of the figure, note that the return track follows the same line as 
the direct one, but it lies in the other half-space with respect to \HH. As long as it 
will be possible, the return track will follow the line 
followed by the direct one. Also note that the direction of the motion is defined
by the order of the decorations in the patter \ftt{234 }{}. It is a reason why we need
a pattern with three elements: with two ones, the orientation would be ambiguous.
Another reason is given by the totalistic constraint: a two-elemented pattern would
also raise ambiguity.

\def\hhzz{\hskip 0.75pt}
\def\sww{{\ftt W }\hhzz}
\def\sss{{\ftt B }\hhzz}
\def\srr{{\ftt R }\hhzz}
\def\sgg{{\ftt G }\hhzz}
The decorations of the elements in the pattern \ftt{234 } are defined by~(3):
\vskip 5pt
\ligne{\hfill
$\vcenter{\hbox{\vtop{\leftskip 0pt\parindent 0pt\hsize=150pt
\ligne{\ftt{2 } $\Rightarrow$ $\Delta_i$, $i$ = 7, 8:\sgg, others:\sww\hfill}
\ligne{\ftt{3 } $\Rightarrow$ $\Delta_i$, $i$ = 7, 8:\sgg, 11:\srr, 
others:\sww\hfill}
\ligne{\ftt{4 } $\Rightarrow$ $\Delta_i$, $i$ =7, 8, 10:\sgg, 11:\srr, others:\sww\hfill}
}}}$
\hfill(\numerrel)\hskip 10pt}
\vskip 5pt

The indices of the neighbours constituting the decoration of an element of the track 
in~(3) are not at all mandatory. The reason is that the new state of the cell depends
on its current state and on the sum of the current states of its neighbours. That sum
does not depend on which faces the milestones are placed. That feature gives us some
freedom in the place of the milestones. The main constraint is that the milestones of
an element~$\eta$ of the track~$\tau$ should not be seen by a neighbour of~$\eta$ also 
belonging to~$\tau$. Also, it should not be seen by the elements of the return track
of~$\tau$. The constraint given in the definition of the return track entails that
an element of the return track of a track~$\tau$ cannot see any element of~$\tau$.
As we consider that, most often, the face~0 of an element of the track lies on~$\tau$,
we may consider that the milestones belong to the upper crown of the element and that 
their place in the crown is most often arbitrary. Of course, we shall mention the right
place when it is needed to fix it.

   The freedom which is given by the totalistic condition gives us a large flexibility
in order to connect two consecutive tracks of a path. We postpone
a more precise description to the next sub-subsection, Sub-subsection~\ref{sssswitch}.

\subsubsection{The switches}\label{sssswitch}

   We define a {\bf path joining} $A$ to~$B$, where $A$ and $B$ are two tiles of the 
dodecagrid, as a sequence \hbox{$\{T_i\}_{i\in\{0..n-1\}}$} of tiles such that
$T_i$ and $T_{i+1}$ can see each other for \hbox{$0\leq i<n$-1} but $T_i$ and $T_j$
cannot see each other if \hbox{$\vert i-j\vert > 1$}. We say that $n$ is the 
length of the just mentioned path joining $A$ to~$B$. We call {\bf distance} from~$A$
to~$B$, denoted by dist$(A,B)$, the shortest length among the lengths of the paths 
joining $A$ to~$B$. Clearly, dist$(A,B)=0$ if and only if $A=B$. Clearly too, that 
distance satisfies the triangular inequality. A circle of radius~$r$ in \HH{} 
around~$T$ in \HH{} is the set of tiles in \HH{} whose distance from~$T$ is~$r$. 
Removing the condition to be in~\HH{} we get a {\bf sphere} of radius~$r$ around~$T$.
A {\bf ball} of radius~$r$ around~$T$ is the set of tiles in all spheres of radius~$\rho$
around~$T$ with $\rho\leq r$. If $T\in~$\HH, the trace in \HH{} of a ball of radius~$r$
around~$T$ is a {\bf disc} of radius~$r$ around~$T$.
In many figures in the Poincar\'e's disc, what we call a {\bf window} focusing on a 
tile~$T$ is a disc of radius~3 around~$T$.

The section follows the implementation described in~\cite{mmarXivh3D3st}. We 
reproduce it here for the reader's convenience. The illustrations of the section show us 
what we call an {\bf idle configuration}. The view given by such pictures is a window 
focusing on what we call the centre of the switch.  An idle configuration is a 
configuration where there is no locomotive within the just defined window. 
Figure~\ref{fstab_fx} shows us three windows focusing on the central tile of a passive
fixed switch: according to what we said in Section~\ref{newrailway}, there is no active 
fixed switch.

\def\hww{\hskip-5pt}
\def\hw{\hskip-4pt}
In the three configurations of Figure~\ref{fstab_fx}, the left-hand side track is the
same. There is just a change in the place of the \sgg-tiles of the decoration of the 
central tile as mentioned in the caption of the figure. In the leftmost picture,
the abutting tracks at the central cell place two {\ftt{3 } \hw}-tiles as neighbours of 
the central tile. 

In the middle picture of the figure, if we continued the right-hand side 
track, it would place a {\ftt{2 } \hw}-pattern as a neighbour of the central cell instead
of a {\ftt{3 } \hw}-one. It is the reason why the right-hand side track is replaced by a 
path which consists of the following tiles: (4)-3$_7$, (4)-7$_6$, (4)-2$_7$, (4)-2, (3)-1.
The number inside parentheses refers to the sector, then the second number is that of the
tile in the sector and the possible lower index~$i$ refers to the $i$-neighbour of the
tile. Note that the just mentioned elements satisfy the conditions put on elements of a
path. Keeping the same notation, we note that (4)-3 is a {\ftt{4 } \hw}-tile, so that 
implementing the {\ftt{234 } \hw}-pattern we get, starting from (4)-3: \ftt{423423 }. 
Accordingly, (3)-1 becomes a {\ftt{3} }-tile and it is a neighbour of the central 
tile which is a {\ftt{4} }-tile, so that the expected pattern is continued. We can 
conclude that the motion we defined in Sub-subsection~\ref{ssstracks} is possible on that
implementation of the passive fixed switch.

\vskip 10pt
\vtop{
\ligne{\hfill
\includegraphics[scale=0.8]{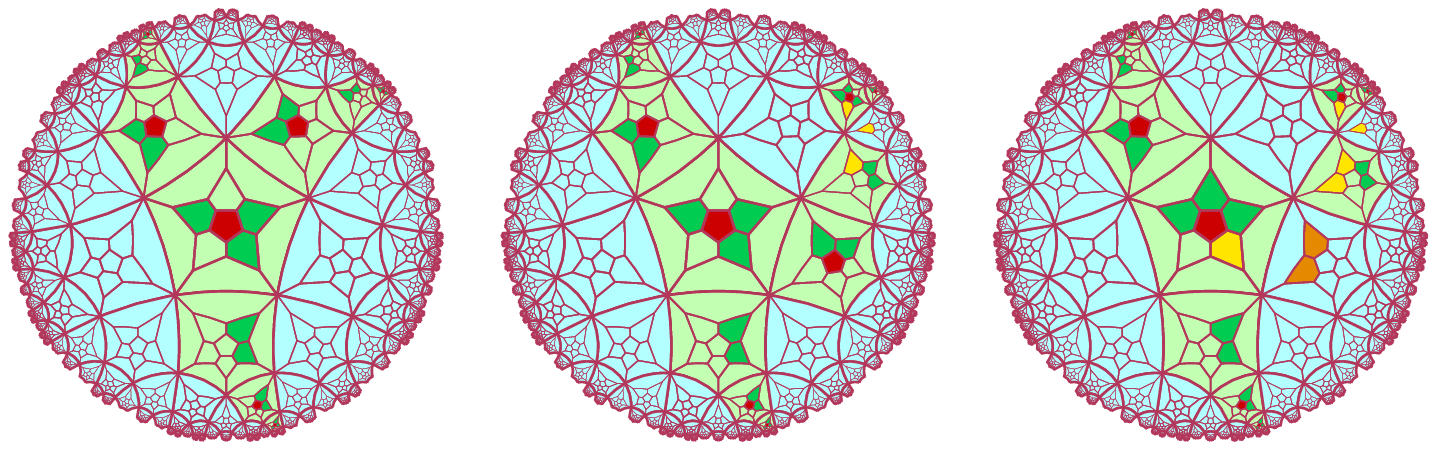}
\hfill}
\vskip-20pt
\begin{fig}\label{fstab_fx}
\leurre
Idle configurations of a passive fixed switch. Note that in the central tile of the
rightmost picture, the place of the {\ftt{4 } \hww}-tiles is not the same as in the 
central tile of the other pictures, but the number of those tiles is the same in the 
three pictures.
\end{fig}
}

In the rightmost picture of Figure~\ref{fstab_fx}, the right-hand side track would abut
the central tile with a {\ftt{4 } \hw}-tile, which does not fit the 
{\ftt{234 } \hw} period. We replace that tile by the following path: 
(4)-3$_7$, (4)-7$_6$, (4)-2$_7$, (4)-2, (4)-2$_6$, (3)-1$_6$, \hbox{(3)-1$_{6,7}$}, 
(3)-1$_7$, 0$_7$. In the 
figure, the tile (3)-1$_{6,7}$ is not represented but its two neighbours (3)-1$_6$ and
(3)-1$_7$ are represented with a different colour as the other tiles of that path
whose elements observe the conditions of an element of a path with respect to its
neighbours in the path. Starting from \ftt{3 } at (4)-3, we get:
{\ftt{3423423423 } \hw}, so that 0$_7$ is a {\ftt{3 } \hw}-tile, which is in agreement 
as a neighbour of~0, as far as 0 is a {\ftt{4 } \hw}-tile.

   Accordingly, we have the solution for implementing the passive fixed switch. We fix
the central tile on a tile~$T$ of a path provided that it is a {\ftt{4 } \hw}-tile.
If the arriving right-hand side branch falls as in the leftmost picture of 
Figure~\ref{fstab_fx}, there is nothing to do. If it is not the case we are then either
in the situation of the middle picture of that figure or in the situation described by
the right-hand side picture. We shall use that solution in other situations.

   We can now turn to the fork, another structure close to the fixed switch. Indeed, we
can see the fork as a reverse passive fixed switch. The totalistic condition allows us to
keep an element of the track as central tile of the fork.

   We decide that the central tile of the fork is a {\ftt{2 } \hw}-tile. Accordingly,
depending on conditions at the other end of the paths leaving the fork, it may be needed 
to tune the arrival at the fork. It is illustrated by Figure~\ref{fstab_frk}
which implements the same solutions as in Figure~\ref{fstab_fx} for the passive fixed
switch.

\vskip 10pt
\vtop{
\ligne{\hfill
\includegraphics[scale=0.8]{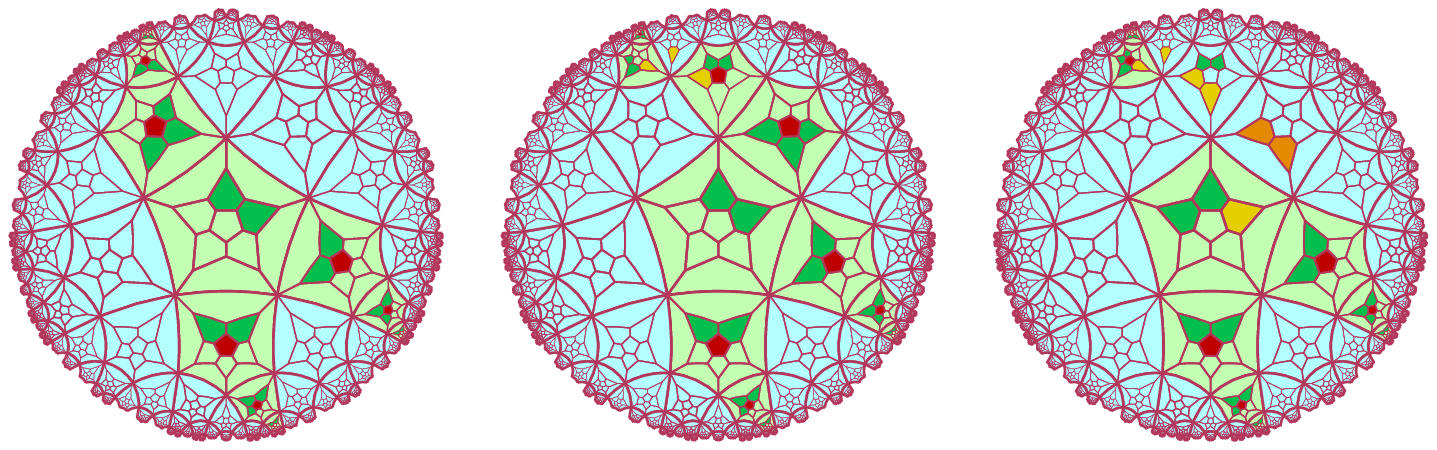}
\hfill}
\vskip-20pt
\begin{fig}\label{fstab_frk}
\leurre
Idle configuration of a fork
\end{fig}
}

   We remain with the implementation of the controller. We postpone its precise 
description to Section~\ref{srules} as far as the controller has a special decoration
as well as the tile which gives access to the controller.

   Before turning to the more complex switches, we have to look at the implementation
of crossings. We take advantage of the flexibility given by the totalistic constraint
for that purpose. Figure~\ref{ftunnel} illustrates the implementation of a tunnel in order
to define a crossing. The figure shows us two tracks, one following a red line, the other
following a blue line. The left-hand side picture is a projection on~\HH{} while the
right-hand side one is a projection on~\VV, a plane which is perpendicular to~\HH{} and
which contains the blue line.

   In the projection on~\VV, we can see that a path joins a tile in~\HH$_u$ to another
one in~\HH$_u$ which is on the other side of the red line. More other, that path does not
contain a neighbour of the central tile, neither by the element themselves nor by the
tiles belonging to the decorations of those elements. We can see that two elements of the
path represented in~\VV{} around the central tile are share a vertex only with the tile
which is represented by the central tile in the projection over~\HH. Not that it could 
also be possible to organise the crossing as a bridge over one of the tracks, but that 
would involve more tiles in the path.

\vskip 10pt
\vtop{
\ligne{\hfill
\includegraphics[scale=0.9]{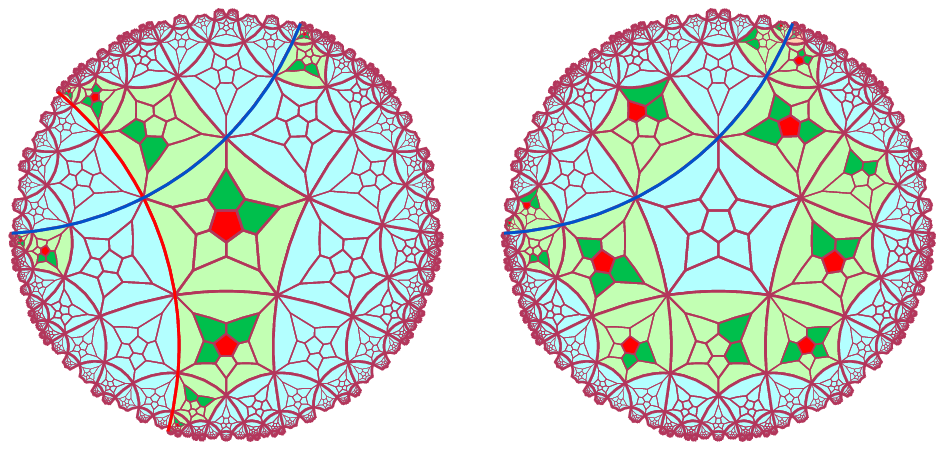}
\hfill}
\vskip-20pt
\begin{fig}\label{ftunnel}
\leurre
Idle configuration of a tunnel. To left, projection on \HH. To right, projection
on \VV.
\end{fig}
}

We are now ready to investigate the flip-flop and the memory switches.

The flip-flop switch and both parts of the memory switch require a much more involved
situation. The global view of an idle configuration of the flip-flop is illustrated by 
Figure~\ref{fschflfl}.

\vskip 10pt
\vtop{
\ligne{\hfill
\includegraphics[scale=0.45]{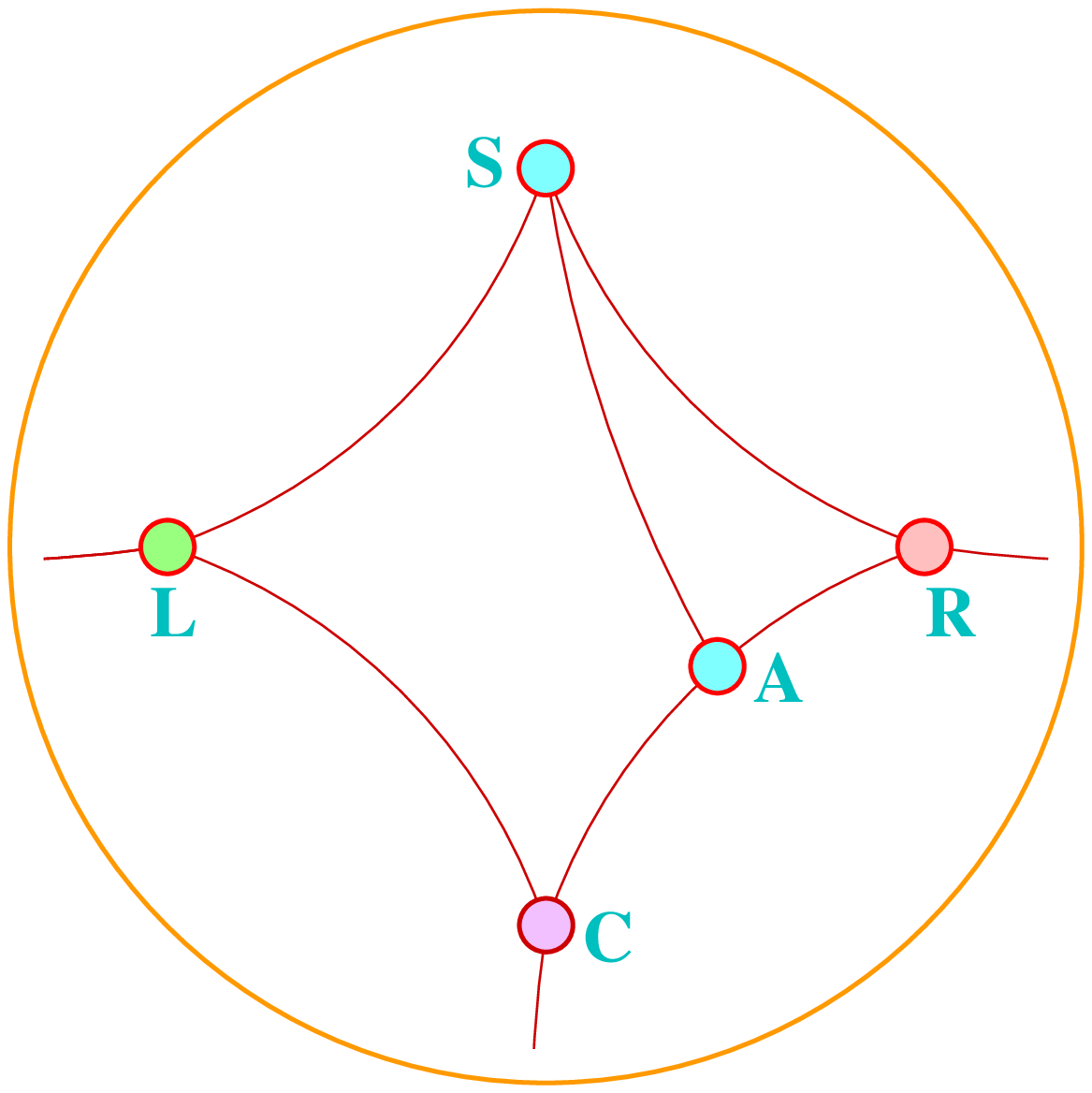}
\raise 50pt\hbox{\includegraphics[scale=0.25]{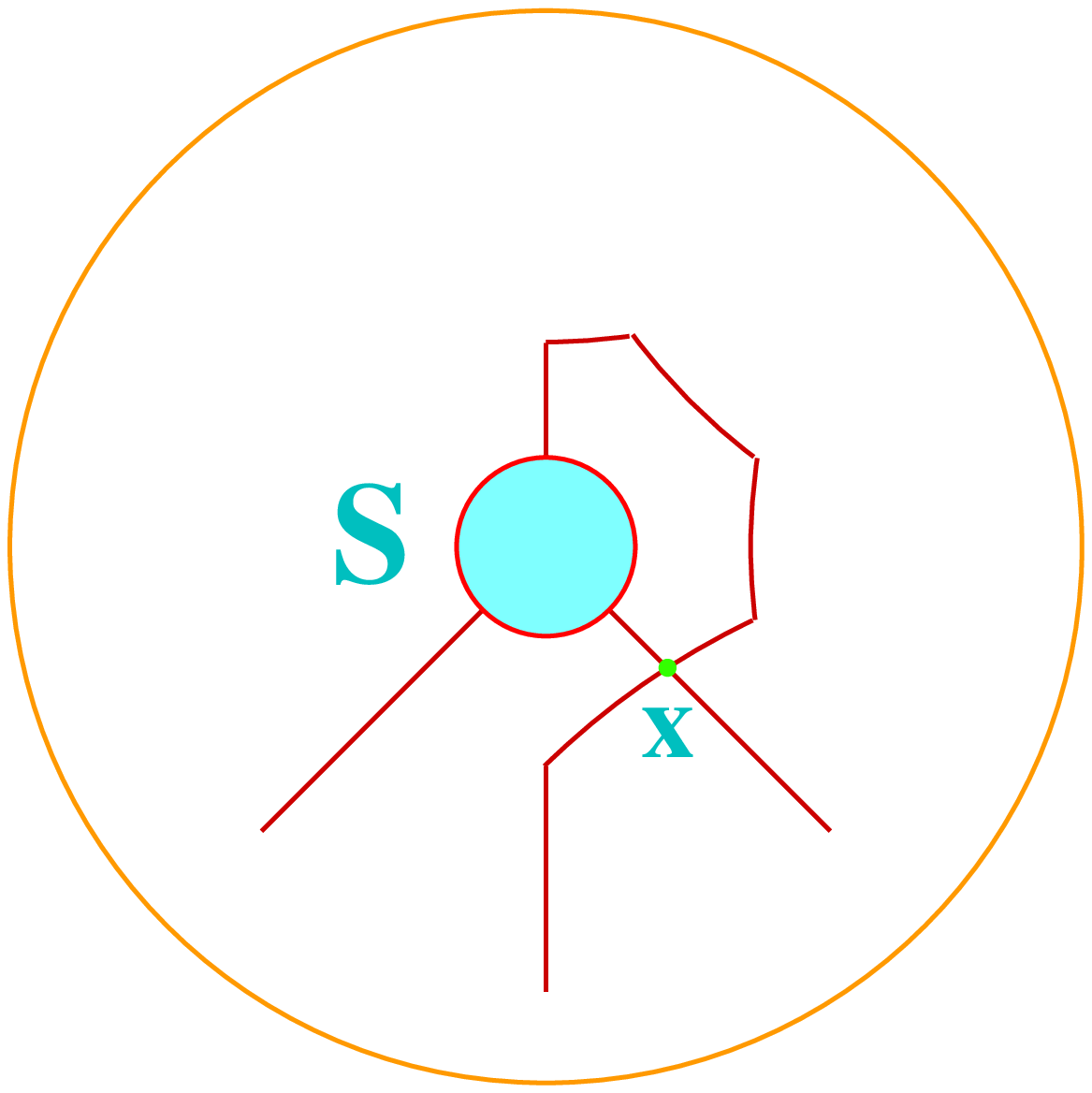}}
\hfill}
\begin{fig}\label{fschflfl}
\leurre
Scheme of the implementation of a flip-flop switch with, to right, a zoom on~{\bf S}.
In the zoom, note the crossing~{\bf x} of a leaving track by the deviated route 
to~{\bf S}.
\end{fig}
}

The locomotive arrives through a segment of a straight line by~$C$ where a fork sits.
Accordingly, two locomotives leave~$C$, one of them towards~$L$, the other towards~$R$.
At~$L$ a controller sits and, on the figure, it let the locomotive go further on the
segment of straight line. Note that the path from~$C$ to~$L$ is also a segment of a
straight line on the figure, which is conformal to the implementation. Now, the structures
which are later involved make the length of that segment to be huge. At~$R$ too a 
controller is sitting but, on the figure, it kills the locomotive which is thus prevented
to go outside the switch. On the way from~$C$ to~$R$ the locomotive meets another fork
at~$A$. The fork sends one of the new locomotives to~$R$ where it is stopped in the 
situation illustrated by the figure and the other is sent to~$S$. There a fork is 
sitting too which sends two locomotives, one of them to~$L$, the other to~$R$. When they
reach their goal, the locomotives change the configuration of the controller which is
sitting there. The controller which let the locomotive go will further stop it while
the one which stopped it will further let it go. Accordingly, after the passage of a 
locomotive and after a certain time, the configuration of the flip-flop switch is that
we described in Sub-section~\ref{railway}.

We have now to clarify the implementation of the controller. It is illustrated by 
Figure~\ref{fctrl}.

Here, the central cell below~\HH{} is not an element of the track. Its decoration
consists of five \sgg-tiles which are placed, for instance on the faces~6, 7, 8, 9 and~10
of the tile. In order to avoid problems with a neighbour, those milestones should be
placed on the upper crown of the tile.

\vskip 10pt
\vtop{
\ligne{\hfill
\includegraphics[scale=1]{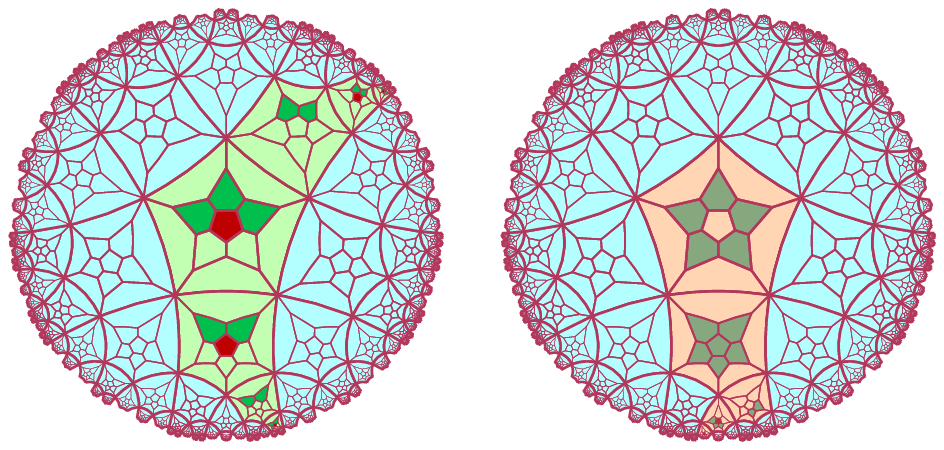}
\hfill}
\begin{fig}\label{fctrl}
\leurre
The idle configuration of the controller. To left, upon~\HH, the track; to right, 
below~\HH, the controller and its access by a locomotive.
\end{fig}
}

   Figure~\ref{fschflfl} requires the explanation of the zoom: three tracks abut the 
point~{\bf S} which is supposed to behave like a fork. However, the configuration of 
the tracks abutting the central tile of a fork is different. It is the reason for which 
the track arriving to~{\bf S} from~{\bf A} is deviated near~{\bf S} as indicated in the
zoom so that the new configuration is conformal to that of a fork. Note that the pieces
constituting the deviation of the track are segments of straight line in Poincar\'e's
disc model. The price to pay is a crossing at~{\bf x} on the picture illustrating the 
zoom. 

\ifnum 1=0 {
   The crossing happens to be a burden in the hyperbolic plane requiring several complex
structures we do not need in the hyperbolic $3D$-space. The third dimension offers
two possible ways to easily realize a crossing: the bridge or the tunnel. In the present
paper, I have chosen the tunnel, illustrated by Figure~\ref{ftunnel}.

\vskip 10pt
\vtop{
\ligne{\hfill
\includegraphics[scale=0.8]{hyp3d_tunnel.ps}
\hfill}
\begin{fig}\label{ftunnel}
\leurre
The idle configuration of the tunnel. 
\end{fig}
}

The leftmost picture of the figure is the standard projection upon~\HH, where each 
dodecahedron is projected within the face it shares with~\HH. In the picture, the faces
of the elements of the track which lie on~\HH{} are faces~0. Using the numbering of the 
tiles of~\HH, the elements of the track going from bottom to top on the picture
are, in this order : sector 3, tiles 21, 8, 3 and~1; the central tile, then, in sector~1,
tiles~1, 4, 12 and 33. Six other tiles can be seen : in sector~2,
tiles 3, 8 and 20 and, in sector~5, 33, 12 and~4 as the track goes from
right to left. Three tiles are missing for that track: tile~1 of sector~5, the central
tile and tile~1 of sector~2. Those tiles are not upon~\HH{} but below that plane.
They are illustrated by the middle picture. It is the part of the crossing
track which passes under the track illustrated by the leftmost picture. In the middle
picture we can see those elements of track as if \HH{} were translucent. Two additional
tiles are indicated in the middle picture: tile~4 of sector~5 and tile~3 of sector~2.
Those tiles correspond to the tiles with the same numbers in the same sectors which lie
upon~\HH. Each of those tiles upon and below~\HH{} allows a locomotive going upon~\HH{}
to go below the plane in order to cross the other track. Those elements require
the locomotive to enter through a face~1 and to exit through a face~2.

\vskip 10pt
\vtop{
\ligne{\hfill
\includegraphics[scale=0.4]{study_tunnel.ps}
\hfill}
\begin{fig}\label{fprojs}
\leurre
Projection of various positions of the element track upon~\HH, \VV{} and also of 
elements below \HH.
\end{fig}
}

The rightmost picture of Figure~\ref{ftunnel} shows us the projection of the tunnel
on the plane~\VV{} which is orthogonal to~\HH, cutting that latter plane along the 
line~$\ell$ which is followed by the three tiles of the tunnel which stand below~\HH.
We represent on~\VV{} the tiles which have a face on it only. Those which are on the
same side of~\VV{} as the central tile in the middle picture are seen in direct
projections. Those which are on the other side are seen as if \VV{} were translucent.

Figure~\ref{fprojs} illustrates the possible projections of an element of the track 
depending on whether it is seen upon~\HH, upon~\VV{} or through those planes by 
transparency, also depending on which face is on~\HH{} and which one on~\VV.

Consider the central tile of the leftmost picture of Figure~\ref{ftunnel}. It is
projected on its face~0 on~\HH{} and the locomotive goes upward, entering through face~5 
and leaving through face~2. Consider the entrance into the tunnel. We can see that
the locomotive enters through face~1 and leave through face~2 on both ends of the tunnel.
According to the role of faces~1 and~2 they occur with face~2 on \HH{} for entering the
tunnel or for leaving it. It is not difficult to see that two faces may be on~\VV{} when
face~2 or face~1 is on~\HH: it is the case for face~0 and face~7. Now, both situations
occur as far as the orientation of the motion is not the same depending on where the
tile are placed. In the first line of Figure~\ref{projs}, we illustrate the projection
of the element track on faces~0, 1, 2, 5 and~7. We choose the case the right-hand side
upper face receives the smallest number. In the second line, we have the same tiles
seen through a translucent face of projection: it is symmetric in the plane which is
perpendicular to the bottom face of the corresponding tile of the first line. 
Now, the tiles we need are rotated images of tiles belonging to the first and to the
second lines. They are given in the third and fourth lines of Figure~\ref{fprojs}.

The first tile on the third line, picture~11, is a rotated image of picture~1 of 
Figure~\ref{fprojs}. It corresponds to the tiles of the path going from bottom to top 
in the leftmost picture of~Figure~\ref{ftunnel}. Picture~12, a rotated image of picture~6
is used in the middle picture of Figure~\ref{ftunnel} to represent the three tiles
constituting the tunnel which are seen through a translucent \HH. Pictures~13, 14 and~15
deal with the entry from the path into the tunnel. Picture~13 is the projection on~\VV{}
of the tile~$e_0$ allowing the locomotive to go from the path to the tunnel while 
picture~14 is the same element projected on its face~2 which lies on~\HH. Pictures~15 
and~16 deal with the tile~$e_1$ below~\HH{} which receives the locomotive coming 
from~$e_0$ and opening the way to the three tiles constituting the tunnel. Picture~15
is a projection on~\VV{} on which is the face~7 of~$e_1$ while picture~16 is the
view of~$e_1$ from a translucent~\HH. Pictures~17 to~20 deal with the other end of the
tunnel when the locomotive goes out from it in order to find the next part of the 
interrupted path. Pictures~17 and~18 illustrate the element~$f_0$ preparing the exit of 
the locomotive from the tunnel. The face~2 of~$f_0$ lies on~\HH. Picture~17 represents
the projection of~$f_0$ on its face~7 which lies on~\VV{} while picture~18 is the
projection of~$f_0$ as seen through a translucent \HH. Pictures~19 and~20 illustrate
the element~$f_1$ which receives the locomotive from~$f_0$ and which lead it to the
continuation of the path through its face~2. That tile has its face~1 on~\HH{} and
its face~0 on~\VV: picture~19, 20 is the projection on~\VV, \HH{} respectively. Note that
tiles~$e_0$ and~$f_1$ are turned in such a way that their exit face are on the right 
direction. For~$e_0$ and~$f_1$, we have to manage the path in such a way that  
their face~6i which is covered by a blue dodecahedron does not belong to a neighbouring 
tile of the path: otherwise it could disturb the motion of the locomotive. That raises 
no problem for~$f_1$ as there is face~7{} in between its faces~2 and~6. For~$e_0$, 
its face~6 is not seen from the neighbour~$\nu$ of~$e_0$ sharing its face~1. The
decoration of an element of the track concerns faces around the face which is opposite
to that which lie on~\HH, so that the $\Delta_6$ of~$e_0$ is not in contact with the path.

   To conclude with the tunnel, the occurrence of a locomotive in the central tile of
the tunnel will not stop a locomotive on the upper way: first, the locomotive never
stops on an element of the track and, more over, in such a crossing there is
a single locomotive in a window around the central point of the crossing: if it is
present on one path, it cannot be present on the other one.
}
\fi

   We are now in position to deal with the memory switch, active and passive parts.

   We first deal with the active part. It looks like the flip-flop switch with this 
difference that there is no fork~$A$ in between the path from the initial fork~$C$
to~$R$, one of the controllers. Figure~\ref{fmemo} illustrates both parts of the 
memory switch: to left, the active part of the switch, to right, its passive
part.

\vskip 10pt
\vtop{
\ligne{\hfill
\includegraphics[scale=0.4]{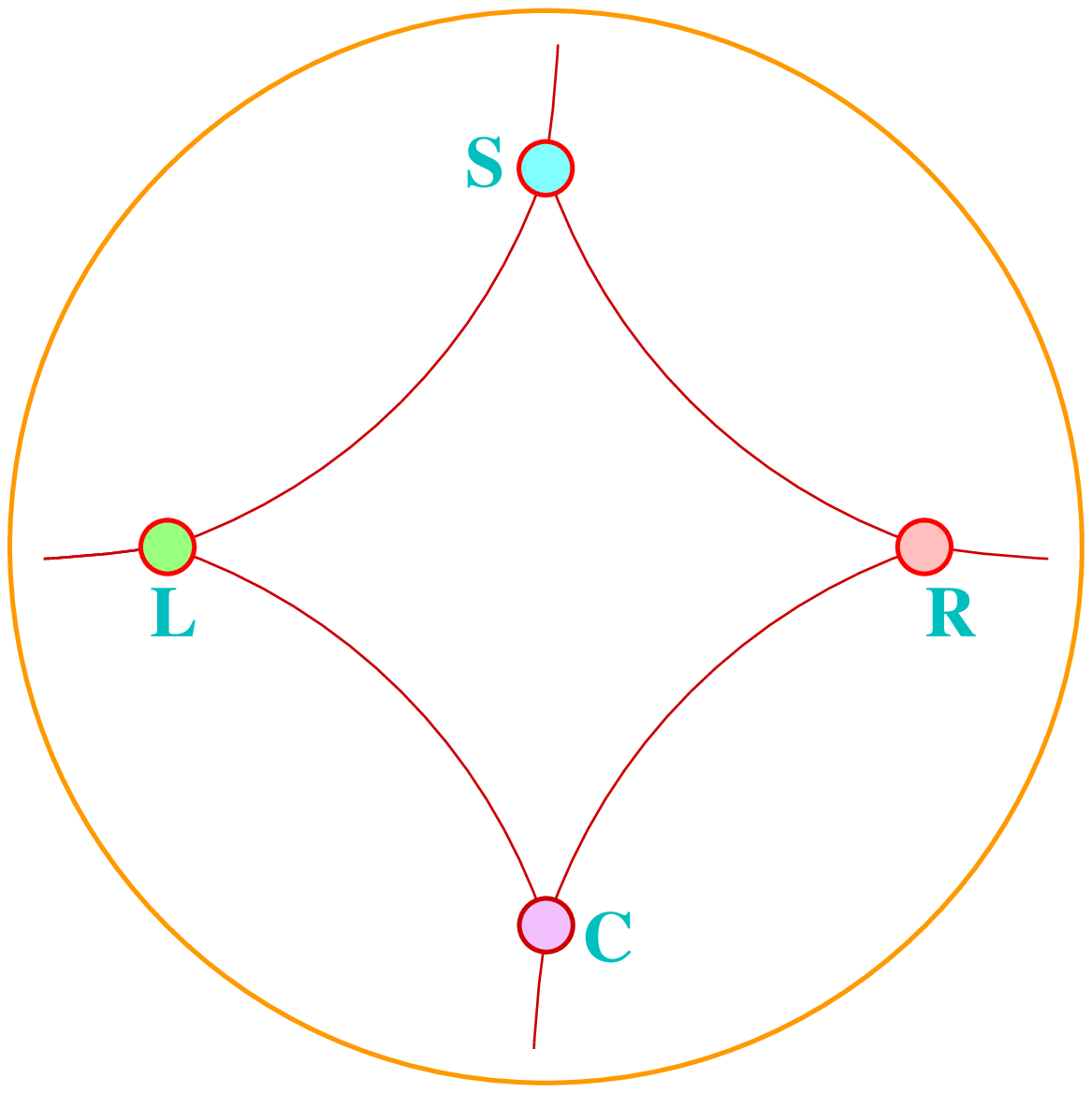}
\includegraphics[scale=0.4]{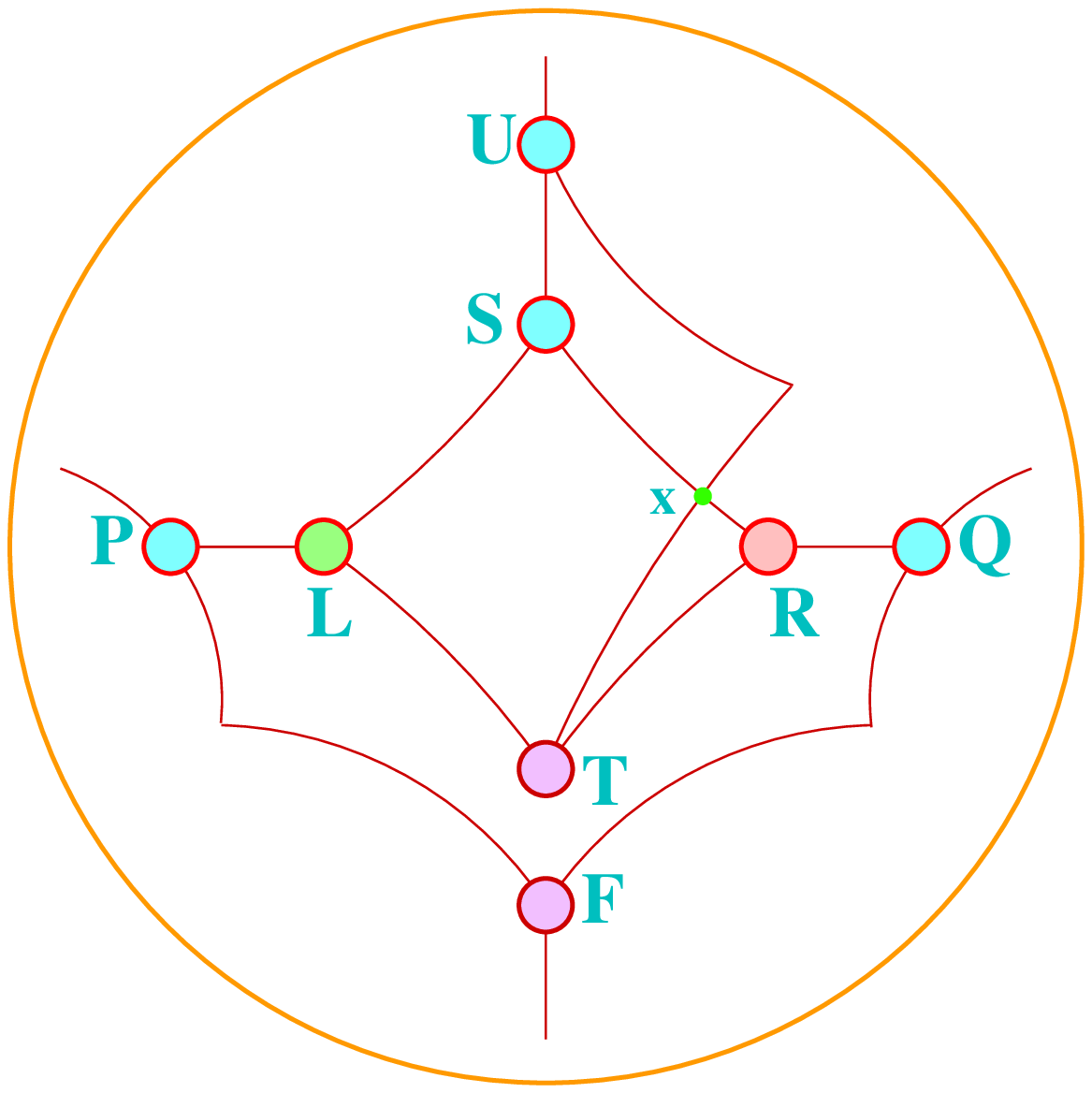}
\hfill}
\begin{fig}\label{fmemo}
\leurre
To left, the active memory switch, to right, the passive one.
\end{fig}
}

In the active part of the switch, left-hand side picture of Figure~\ref{fmemo}, 
the locomotive arrives to a fork sitting at~$C$. From there two locomotives are sent,
one to~$L$, the other to~$R$ and the working of the switch at this point is alike that
of a flip-flop switch. The difference lies in the fact that the passage of the locomotive
does not trigger the exchange of the roles between the controllers. That change is 
triggered by the passage of a locomotive through the non-selected track of the passive
part of the switch. When it is the case, a locomotive is sent from the passive part
to the active one. That locomotive arrives at the fork which is sitting at~$S$. The fork
creates two locomotives which are sent to~$L$ and~$R$ in order to change the permissive
controller to a blocking one and to change the blocking one into a permissive one.

Let us look at the working of the passive part. The locomotive arrives to the switch 
through~$P$ or through~$Q$. Assume as it through~$P$. A fork sitting at~$P$ sends a
locomotive to the fixed switch~$F$ which let the locomotive leave the switch. The other
locomotive sent by~$P$ goes to~$L$. If that side is that of the selected track, the
controller sitting at~$L$ blocks the locomotive so that no change is performed, neither in
the passive switch, nor in the active one. Accordingly,the selected track of the active
switch is controlled by a permissive controller while the corresponding selected track
of the passive switch is controlled by a blocking controller. Presently, assume that
the side of~$L$ is not that of the selected track. It means that $L$ let the locomotive
go to~$T$ where a fixed switch sends the locomotive to a fork at~$U$. That fork
sends a locomotive to the fork~$S$ of the active switch and the other locomotive is sent
to~$S$ of the passive switch. At that point~$S$ in the passive switch, a fork sends a 
locomotive to~$L$ and another one to~$R$ in order to change the working of both 
controllers to the opposite task. As a parallel change occurs in the active switch the 
selected track is redefined in both parts of the switch.

   That working raises several remarks. First, the role of the controllers in the active
and in the memory switches are opposite. Nevertheless, in both cases, the same
programmable controller is used exactly because it is programmable in the way we just
described. A second remark is that we used one crossing, two
fixed switches, at $F$ and at~$T$ and four forks, at~$P$, $Q$, $U$ and~$S$. Each
structure requires some space, at least a disc whose radius is the length of four tiles
aligned along a straight line. Consequently, the passive memory switch requires a huge
amount of tiles.

\subsubsection{The one-bit memory}\label{sssunit}

It is now time to implement the one-bit memory. Figure~\ref{fonebit} illustrates the
construction.

\def\WW{{\bf W}}
\def\DD{{\bf D}}
\def\RR{{\bf R}}
\def\EE{{\bf E}}
\def\ZZ{{\bf Z}}
\def\bbz{{\bf b0}}
\def\bbu{{\bf b1}}
\def\zz{{\bf 0}}
\def\uu{{\bf 1}}
We can see the active memory switch at~$R$ and the passive one at~$E$. The dark lettres
which stand by the blue circle indicate {\bf gates} of the one-bit memory: \WW, \RR, \EE,
\bbz{} and \bbu. We can easily see that if the locomotive enters the unit through 
the gate~\RR, then it leaves the memory through the gate~\bbz{} or through the gate~\bbu{}
depending on the information stored in the memory: that information is provided the unit
by the positions of the switch at~\WW{} and those at~\RR{} and~\EE. Note that the
positions at~\RR{} and at~\EE{} are connected by the path from~\EE{} to~\RR, see the
figure.

\vskip 10pt
\vtop{
\ligne{\hfill
\includegraphics[scale=0.5]{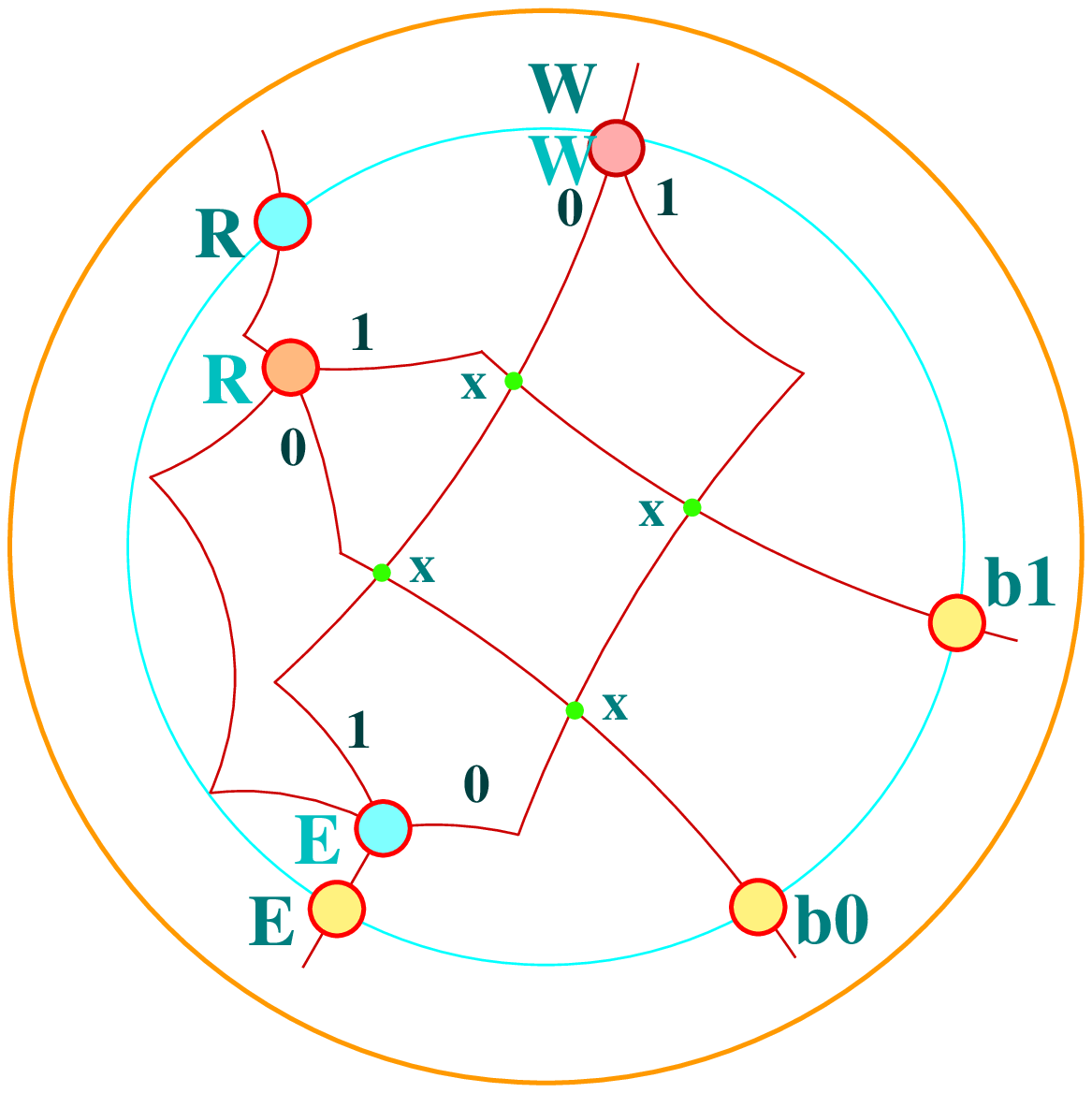}
\hfill}
\begin{fig}\label{fonebit}
\leurre
The idle configuration of the one-bit memory. Note the four crossings in the 
implementation. Note that the connection from~\EE{} to~\RR{} is realized by three 
segments of straight lines.
\end{fig}
}

When the locomotive enters the memory through the gate~\WW{} where a flip-flop switch is
sitting, it goes to~\RR{} through one of both tracks leaving the switch. If it goes
through the track marked by~\zz, \uu, it arrives to~\EE{} by the track marked with the
opposite symbol, \uu, \zz{} respectively. Indeed, when the locomotive crosses~\WW,
the passing makes the selected track to be changed so, if it went through one track, after
the passage, in particular when the locomotive arrives at~\EE, the new selected track
at~\WW{} is the track through which the locomotive did not pass. So that the
track marked by one symbol at~\WW{} should be marked by the opposite one at~\EE. 

   As the one-bit memory will be used later, we introduce a simplified notation:
in Figure~\ref{fonebit}, the memory structure is enclosed in a blue circle. At its 
circumference the gates are repeated by the same symbols. In the next figures, when a
one-bit memory will be used, we shall indicate it by a light blue disc with, at its 
border, the five gates mentioned in Figure~\ref{fonebit}.

\subsubsection{From instructions to registers and back}\label{sssregdisp}

   As will be explained in Sub-subsection~\ref{sssreg}, the locomotive arrives at a 
register at a point which depends on the type of the operation to be performed. It
depends on the type only, whether it is a decrementation or an incrementation. It does
not depend on namely which instruction of the program required the execution of that
operation. Moreover, the return path of the locomotive once it performed its operation
is the same in most cases. Accordingly, when the locomotive goes back from the register
to the program, it is important to define the point at which it will return. 
Correspondingly with what we said, the unique solution is to keep track of that
information before the locomotive enters the register where that information disappears.

\def\DDI{${\mathbb D}_I$}
\def\DDD{${\mathbb D}_D$}
\def\DDO{${\mathbb D}_O$}
   To that goal we define a structure~\DDI{} for each register as follows. The structure 
consists of as many units as there are instructions incrementing that register, say $R$, 
in the program. Each unit is based on a one-bit memory and Figure~\ref{fdispinc} 
illustrates such a unit. The working of~\DDI{} is the following. An instruction for 
incrementing $R$ is connected through a path to a specific unit of~\DDI. The path goes 
from the program to the \WW-gate of that unit. At the initial time, the configuration 
of~\DDI{} is such that all its unit contain the bit~0: the switches of the one-bit memory 
are in a position which, by definition defines bit~0. Accordingly, when the locomotive 
enters the unit, it will change the flip-flop and the memory switches so that, 
by definition, the memory contains the bit~1. The locomotive leaves the memory through 
the gate~\EE{} and it meets a flip-flop switch at~$A$ which, in its initial position, 
sends the locomotive to~$R$. Note that when the locomotive leaves~\DDI{} a single unit
of the structure contains the bit~1 and selects a path to the program at its flip-flop
switch.

Similarly, a structure \DDD{} memorises which instruction required a decrementation
of that register. The structure contains as many unit as there are instructions of
the program which decrements that register. But instead of containing one bit, each unit
contains two of them. Those bits are either both~0 or both~1. Before the arrival of a 
locomotive to~\DDD, both bits of each unit are set to~0. When the locomotive enters
the structure, it arrives at the unit which corresponds to the instruction which sent it
to that register. The locomotive sets both bits of the unit to~1. When it leaves~\DDD,
a single unit contains its both bits set to~1.

But just before entering the register, the locomotive enters a third structure, \DDO,
which memorises which type of instruction operates on the register. The structure
contains two one-bit memories which are set to~0{} in its initial configuration. If the 
locomotive comes from~\DDI{} both bits remain~0. When it comes from~\DDD, both bits 
are set to~1.

Accordingly, when the locomotive goes back from the register, if the bits of~\DDO{}
are set to~0, it goes back to~\DDI. There, the locomotive goes to the unique unit
whose bit is set to~1 and the unit sends the locomotive back to the program at the
right instruction by an appropriate path. If the bits of~\DDO{} are set to~1,
the arrival to~\DDO{} depends on whether the decrementation was possible or not: there
are two arriving paths to~\DDO. The locomotive reset both bits to~0 and it leaves
the structure through the appropriate path: that which corresponds to a successful
decrementation or that which corresponds to the case of a register containing~0.

Let us first consider the case of~\DDI. The structure of a unit is illustrated by
Figure~\ref{fdispinc}.

When the locomotive completed the incrementation of~$R$ it goes back to the program.
In order to find the appropriate way, it visits the units of the~$D_I$ attached to~$R$
until it finds the single unit whose one-bit memory is set to~1. Then, the locomotive 
rewrites the bit, turning it to~0 and again exits through the gate~\EE. It meets
the flip-flop switch at~$A$ which sends the locomotive on its other path which leads it 
back to the program. That new visit of the flip-flop switch at~$A$ makes the switch 
again select the track leading to~$R$. Accordingly, when the locomotive leaves~\DDI{} 
the structure recovers its initial configuration.
\vskip 10pt
\vtop{
\ligne{\hfill
\includegraphics[scale=0.5]{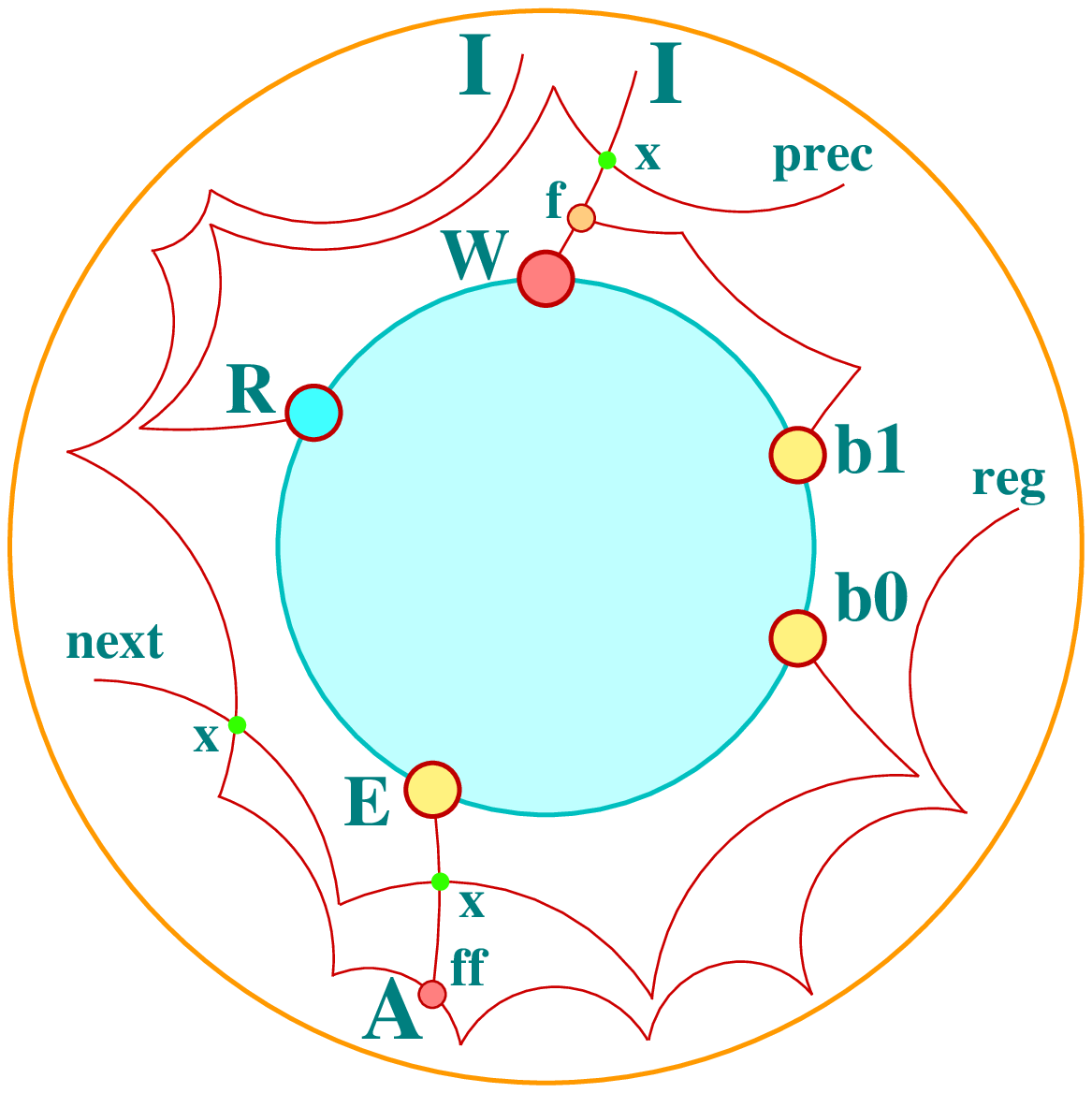}
\hfill}
\begin{fig}\label{fdispinc}
\leurre
The idle configuration of a unit of the structure which memorises the right incrementing
instruction. Note the three crossings in the 
implementation. Also note that the intensive use of sequences of connected segments of
straight lines.
\end{fig}
}

We can check on Figure~\ref{fdispinc} that the implementation it gives for a unit
entails a working which is conformal to the above scheme. 

When the locomotive comes from an incrementing instruction of the program, it arrives
at the corresponding unit of the right~\DDI{} through the gate~\WW{} of that unit.
Accordingly, it changes a bit~0 to~1 and leaves the memory through~\EE. From there it 
meets~$A$ where the flip-flop switch sends it to~$R$: as the unit contained~0, the switch 
at~$A$ selects the path leading to~$R$. When it leaves~$A$, the flip-flop now selects the
other path, that which leads back to the program. 

When the locomotive comes back from~\DDO, it arrives to the closest unit of~\DDI{}
through its gate~\RR. If the locomotive reads~0, it leaves the unit through~\bbz{}
and the corresponding path sends the locomotive to the next unit.
Accordingly, the locomotive crosses the units containing~0 until it arrives to
the \RR-gate of the single unit containing~1. As far as it reads~1, the locomotive leaves 
the memory through~\bbu{} on a path which leads it to~\WW{} again.
Accordingly, the locomotive rewrites the content of the memory which is thus changed
from~1 to~0. The locomotive again leaves the memory through~\EE{} and meet~$A$
where the flip-flop switch presently sends it to the program. When leaving $A$, the
switch is changed and it again selects the path leading to~$R$, so that that \DDI{}
recovers its initial configuration. The scheme allows us to correctly simulate the 
working of the units of~\DDI.

Presently, we examine the structure \DDD{} which plays for the 
decrementing instruction the role that \DDI{} does for the incrementing ones. The 
structure is more complex for the following reason. An incrementing instruction is 
always performed
which is not necessarily the case for a decrementing instruction. Indeed, if the register
contains the value~0, it cannot be decremented. We say that the register is empty. In 
that case, the next instruction to be performed is not the next one in the program. It 
is the reason why the case of an empty register requires a particular treatment. 
Concretely, it means that the return track of the locomotive depends on whether the
register was empty or not at the arrival of the locomotive.

\vskip 10pt
\vtop{
\ligne{\hfill
\includegraphics[scale=0.45]{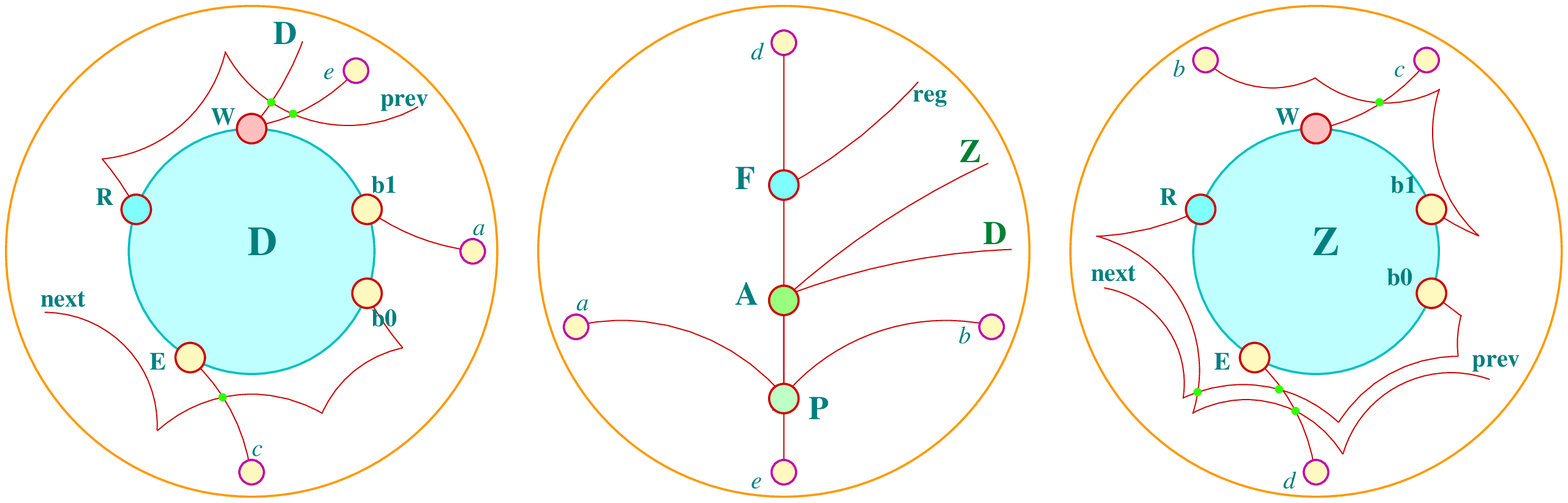}
\hfill}
\begin{fig}\label{fdispdec}
\leurre
The idle configuration of a unit of the structure which memorises the right decrementing
instruction. As far as the structure is complex, its components are displayed in three
windows. The yellow small discs close to the border of a circle indicate points of
communication between the windows: the track arriving to a small disc with an italic 
small lettre is continued by the track issued from a small disc with the same italic
lettre. Note the sketchy representation of the one-bit memories.
\end{fig}
}

Accordingly we need two one-bit memories instead of one. More over, as far as the bit~1
locates the unit of~\DDD{} which will send the locomotive to the right place in the 
program, the content of the memories should be the same: 0, if the locomotive did not
visited that unit at its arrival to~\DDD, 1 if it visited the required unit. As
there are two return tracks after decrementation, the $D$-track is 
used when the decrementation could be performed, the $Z$-track is used when the 
decrementing locomotive found an empty register. We call \DD-, \ZZ-{\bf memory} 
the one-bit memory visited by $D$-, $Z$-track respectively. The complication entailed by 
that organisation makes a representation of a unit in a single window hardly readable.
It is the reason why it was split into three windows as illustrated by 
Figure~\ref{fdispdec}.

   To left, we have the \DD-memory. It is represented by the left-hand side picture of
the figure. A locomotive sent by the program for decrementing a register~$R$ arrives
at the appropriate unit of the \DDD{} attached to~$R$ by the track marked by $D$ in the 
picture. As far as the locomotive enters the memory through its \WW-gate, it rewrites 
its content
from~0 to~1. Now, it has to mark the \ZZ-memory which is represented on the right-hand
side picture of the figure. To that goal, the track leaving the $D$-memory through its
gate~\EE, goes to the point~$c$, that italic lettre marking a small yellow disc close to
the border of the window. The track is continued from a small yellow disc close to the 
border of the right-hand side window marked with the same lettre~$c$. Other such discs
in the figure are marked by other lettres which are pairwise the same in two different 
windows. From that second $c$, the track leads the locomotive to the gate~\WW{} of
the \ZZ-memory whose content, accordingly, will be changed from~0 to~1. Leaving the
\ZZ-memory through its gate~\EE, the locomotive is lead to a small disc~$d$ so that the
corresponding track is continued by the small disc~$d$ we find in the middle window of
Figure~\ref{fdispdec}. From there, the locomotive is lead to a flip-flop switch sitting 
at~$F$ which, in its initial configuration, selects the track leading to~$R$. Now,
after the switch is passed by the locomotive, its selection is changed, indicating the
track leading to another switch sitting at~$A$.

   Consider the case when the locomotive returns from~$R$ after a successful 
decrementation. It one by one visits the units of~\DDD. It arrives to the gate~\RR{}
of the \DD-memory of a unit. If it reads~0, the track issued from~\bbz{} of the 
\DD-memory leads the locomotive to the \RR-gate of the \DD-memory of the next unit.
So that such a motion is repeated until the locomotive reads~1{} in the unique \DD-memory
whose bit is~1.
When it is the case, the locomotive leaves the memory through its \bbu-gate which
leads the locomotive to a small disc~$a$, so that the track is continued from the
small disc~$a$ we can see in the middle picture of Figure~\ref{fdispdec}. From there,
the locomotive arrives to~$P$, the passive part of the memory switch whose active part
lies at~$A$. The locomotive is sent from~$P$ to a small disc~$e$ whose track arriving
there is continued by the track issued from the small disc~$e$ of the left-hand side
picture of the figure. That track leads the locomotive back to the \WW-gate of the
\DD-memory. Consequently, the content~1 of the memory is returned to~0 and leaving
the memory through its gate~\EE, the locomotive again arrives to the \WW-gate of the
\ZZ-memory through the small discs~$c$. That visit rewrites the bit of the \ZZ-memory 
from~0 to~1. Then, the locomotive leaves the \ZZ-memory through its \EE-gate which sends
the locomotive to the small disc~$d$ which, through the other such disc in the middle
window, again arrives at $A$. As far as it took much more time for the locomotive to 
rewrite both memories of the unit than for another locomotive to go from~$P$ to~$A$
in order to set the selection of the active switch, the switch a~$A$ indicates the
track corresponding to the branch $P$$a$ of the passive switch. That track leads to
the right decrementing instruction in the program.

   Presently, consider the case when the locomotive returns from an empty $R$.
It returned through the \ZZ-track and it arrives to the~\DDD{} attached to~$R$.
There it visits the units of the structure reading the content of its \ZZ-memory through
the track leading to the \RR-gate of that memory in the unit. So that the locomotive 
visits the units until the single one whose \ZZ-memory contains~1. When it is the case,
the locomotive leaves the \ZZ-memory through its \bbu-gate which sends it to the 
\WW-gate of the \DD-memory through the small disc~$b$. But the continuation happens by~$b$
of the middle picture which leads the locomotive to~$P$. The passive switch sends the 
locomotive to~$e$ and from there to the \WW-gate of the $D$-memory, $P$ remembering that
the locomotive arrived through the $P$$b$-branch of the switch. Accordingly, what
we seen in a visit through a unit thanks to the $D$-track occurs once again. 
Accordingly, the locomotive
rewrites the contents of both the \DD- and the \ZZ-memories, returning them from~1 to~0.
Then, the locomotive leaves the \ZZ-memory through its \EE-gate so that, via the disc~$d$,
it arrives to~$F$ where the flip-flop switch sitting there sends it to the switch
sitting at~$A$. Now, during its trip for the \EE-gate of the \ZZ-memory to the \WW-gate
of the \DD-memory, the locomotive arrived to~$P$ through the branch $P$$b$ of that
passive switch. Accordingly, the switch selected that track and, in the meanwhile, sent
another locomotive to~$A$ in order to select the appropriate track, that which leads
to $Z$ in the program. As the locomotive passed for a second time through~$F$, the
switch selects presently the track leading to~$R$, its initial configuration, up
to the configuration at~$A$ which can be arbitrary as shown by our discussion. 
Accordingly, the scheme illustrated by the picture performs what is expected.

   Presently, we arrive to the structure~\DDO{} which is illustrated by 
Figure~\ref{fdispop}. As mentioned in the caption, the figure very much looks like
Figure~\ref{fdispdec}. Indeed,the middle picture is almost the same in both figures while
the other pictures are a bit simpler in Figure~\ref{fdispop}. Indeed, in the middle 
picture, the paths leaving~$A$ go to different places that in Figure~\ref{fdispdec}.

\vskip 10pt
\vtop{
\ligne{\hfill
\includegraphics[scale=0.45]{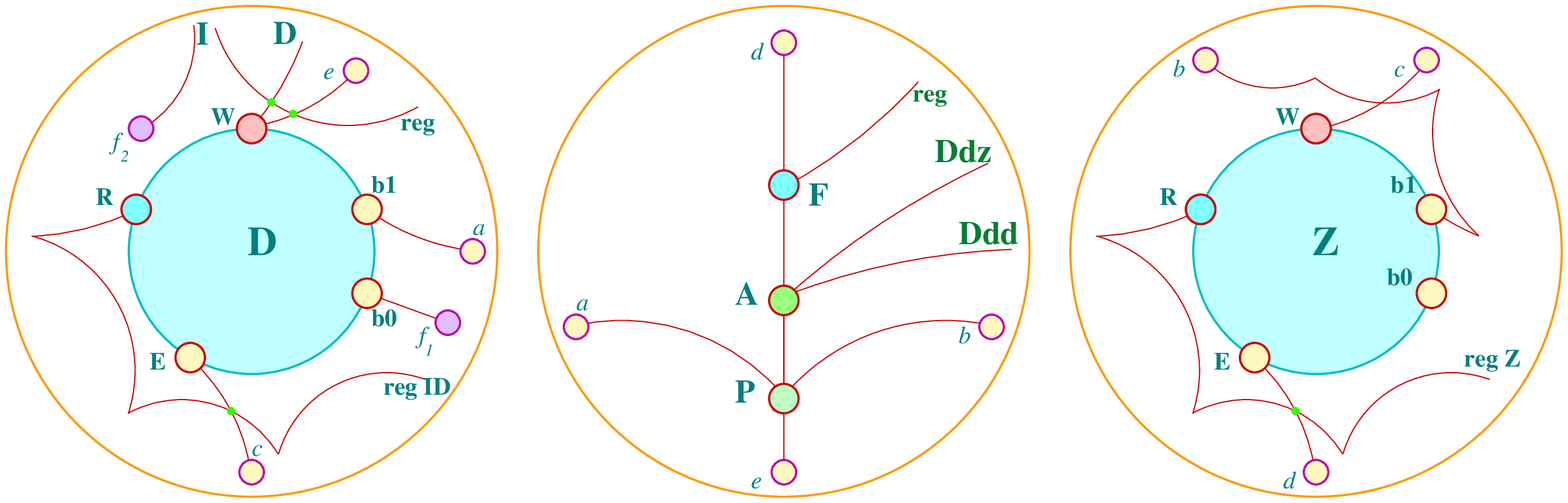}
\hfill}
\begin{fig}\label{fdispop}
\leurre
The idle configuration of a unit of the structure which memorises the type of
the current instruction. The structure looks like very much a unit of \DDD{}
as displayed by Figure~{\rm\ref{fdispinc}} although it is a bit simpler.
\end{fig}
}

   As indicated by the leftmost picture of Figure~\ref{fdispop}, when the locomotive
comes from an incrementing instruction via a \DDI-structure, it does not cross the 
\DDO-structure which remains with both its memory-units set to~0. It directly goes
to the register via a path which is specific to incrementing instructions for that
register. As the one-bit memories will be distinguished in what follows, we denote them 
by \DD{} and \ZZ{} respectively as in a unit of a \DDD-structure.  When the locomotive 
comes from a decrementing instruction via a \DDD-structure, it enters the \DDO{} of the 
register through the \WW-gate of its \DD-memory. Accordingly, the bit of that unit is 
set to~1. The locomotive exits from the \DD-memory through the gate \EE{} and then, 
through the small circle~$c$, it is sent to the \WW-gate of the \ZZ-memory, see the 
small disc $c$ in the rightmost picture of Figure~\ref{fdispop}. Accordingly, the bit 
of the \ZZ-memory is set to~1. Exiting from
the gate \EE{} of the \ZZ-memory, the locomotive is sent via the discs~$d$ to that of the
middle picture of the figure. From that $d$-disc, the locomotive is sent to a flip-flop
switch sitting at~$F$ which sends the locomotive to the register, through a path which is
specific to the decrementing instructions for that register. After crossing~$F$, the
switch sitting there now selects the track leading to the switch at~$A$.
   
   Consider the case when the locomotive returns from the register after executing
an instruction. It is the same track if the instruction could be completely performed.
The track arrives at the gate~\RR{} of the \DD-memory. It the locomotive reads~0,
the \bbz-gate sends it to the program, to the instruction following the incrementing 
instruction which was executed. The \bbz-gate sends the locomotive to the small 
disc~$f_1$, so that it directly goes to the disc~$f_2$ from which the track leads to the
program. If the locomotive reads~1, the \bbu-gate sends the locomotive to a disc~$a$ 
which indicates the continuation through the disc~$a$ of the middle picture of the figure.
There, the locomotive is sent to the passive memory switch sitting at~$P$. Accordingly,
that switch orders the switch at~$A$ to select the path leading to the \DDD{} of the
register. After crossing~$P$, the locomotive arrives at the small disc marked by~$e$,
so that it continues its way through the small disc~$e$ of the leftmost picture of the
figure. Then the enters the \DD-memory through its \WW-gate, so that it rewrites the 
bit from~1 to~0. Leaving the memory through~\EE, the locomotive is sent through a small 
discs~$c$ to the \WW-gate of the \ZZ-memory whose bit is turned from~1 to~0. The
locomotive is sent through the small discs~$d$ to the flip-flop switch at~$F$. As
the \DDO-structure had its memories set to~1, the switch at~$F$ indicates the path to~$A$.
Now, when the locomotive had set the \DD-memory back to~0, it arrived to~$P$ through~$a$
so that the switch at~$A$ indicates the path to~\DDD. As far as the flip-flop switch 
at~$F$ was crossed a second time, it now indicates the path to the register. Accordingly,
when the locomotive leaves the \DDO{} of the register, that structure recovers its
initial configuration up to the switch at~$A$ whose selection is arbitrary.

We remain with the case when a decrementing instruction failed to decrement a register
as far as that one was empty. The return track is not the same as previously. It is
a specific track, the $Z$-path, which also visits the \DDO-structure: that structure has 
its one-bit memories set to~1, they must be returned to~0. Now, the locomotive arrives 
through the $Z$-path at the \RR-gate of the \ZZ-memory. As far as the decrementing
instruction visited \DDO{} before arriving at the empty register, the \ZZ-memory bit
contains~1, so that the locomotive leaves that memory through its \bbu-gate which
sends the locomotive to the small disc~$b$. The continuation happens in the middle picture
where the small disc marked with~$b$ sends the locomotive to~$P$ where a passive memory
switch sits. As far as the locomotive arrives at~$P$ through the branch $Pb$ of the 
switch, $P$ makes the active switch sitting at~$A$ select the track leading to \DDD{}
via the $Z$-path. The locomotive leaves $P$, going to~$e$ where the continuation occurs
in the leftmost picture of the figure. There, the locomotive is sent to the \WW-gate of
the \DD-memory whose bit is accordingly turned from~1 to~0. As in the case we have
seen f a successful decrementation, the locomotive via the small discs~$c$ is sent to
the \WW-gate of the \ZZ-memory whose bit too is changed from~1 to~0. At~$F$, the flip-flop
switch indicates the path to~$A$ as far as when it arrived to~\DDO{} the bits of the
structure were set to~1, so that the locomotive is sent to \DDD{} via the $Z$-path.
As far as the locomotive crossed~$F$ the flip-flop switch changes its selection so
that it again selects the path to the register. Accordingly, the locomotive leaves~\DDO{}
in the initial condition of the structure, up to the switch at~$A$ whose selection is
arbitrary.

   At last and not the least, we have to look at what happens when the locomotive arrives
to the program after performing its operation on the register. Two cases occur: the
new instruction is the next one in the program after the previous one; the new 
instruction occurs at another place in the program. In both cases, the return track leads
the locomotive to the appropriate place.

\subsubsection{Constitution of a register}\label{sssreg}

   The implementation of the register requires a special examination. Weak universality
means that the initial configuration is infinite but not arbitrary. In the present paper,
it will be periodic outside a large ball containing the implementation part of the
program and also the first unit of the two registers needed for universality, according
to Minsky's theorem, see~\cite{minsky}. Each register, in some sense, follows a line
and that construction along each line is periodic.

   A register consists of infinitely many units which we may index by $\mathbb N$.
Let $\mathcal R$ denote a register. By $\mathcal R$($n$), we denote the $n^{\rm th}$
unit. We shall call $\mathcal R$(0) the first unit of the register.
Each unit contains two one-bit memories. Both memories contain the same bit when the 
locomotive is away from the register. At each time~$t$ of the computation, there is
a number~$c_t$ such that the bits of $\mathcal R$($n$) are both set to~0 when
\hbox{$n\geq c_t$} and all of them are set to~1 when \hbox{$n<c_t$}. We say that
$c_t$ is the {\bf value} of the register. We also say that it is its {\bf content}.
When $c_t=0$ we also say that the register is empty. In that case, the bit in all 
memories of all units of the register is set to~0.

\vskip 10pt
\vtop{
\ligne{\hfill
\includegraphics[scale=0.65]{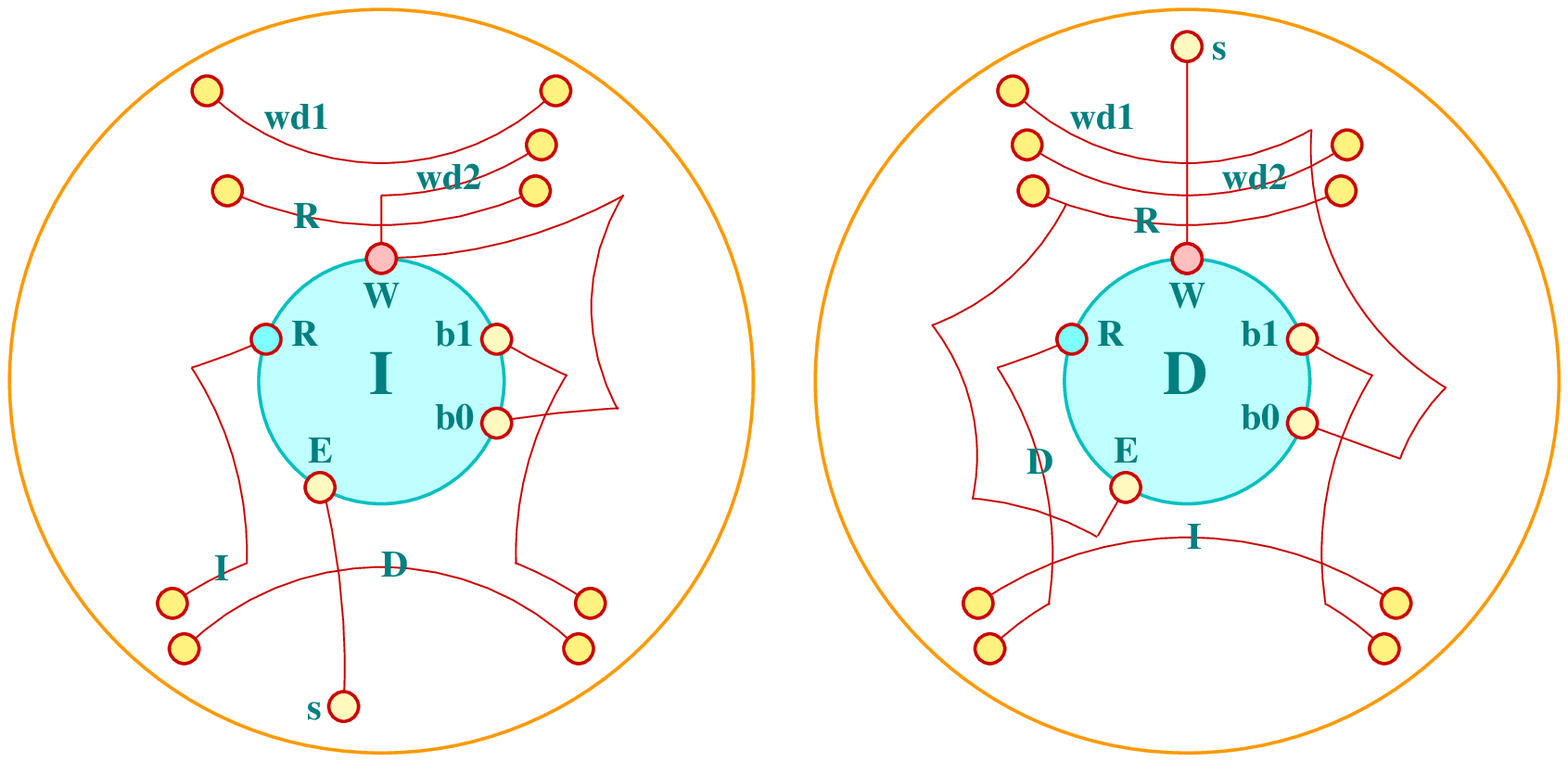}
\hfill}
\begin{fig}\label{funitreg}
\leurre
The idle configuration of a unit of a register. We can see the memories devoted to
the materialisation of the content of the register. On the bottom, the access to the
unit, on the top the return path and the auxiliary paths used for performing a
decrementation.
\end{fig}
}

\def\II{{\bf I}}
On Figure~\ref{funitreg}, we display such a unit. In order to make the structure more 
visible, we use two windows: one focuses on what we call the \II-memory the other focuses
on the \DD-memory. On the lower part of each window, we can see two paths delimited
by small circles. Those small circles are devoted to allow the locomotive to go from
the left-hand side path to the right-hand side one when the paths are marked by the same
lettre. A similar organisation can be seen above the one-bit memories but, this time,
the locomotive goes from the right-hand side window to the left-hand side one. 

\subsection{Operations on a register}\label{ssopreg}

Presently, we describe how the operations on a register are performed by a locomotive.
Sub-subsection~\ref{sssinc} deals with incrementation while decrementation is dealt with
by Sub-subsection~\ref{sssdec}. In both cases of our discussion, the reader is invited
to follow the arguments on Figure~\ref{funitreg}.

\subsubsection{Incrementation}\label{sssinc}

\def\RRR{$\mathcal R$}
When a locomotive arrives from the \DDO-structure of the register \RRR{} at its first 
unit in order to increment \RRR, the locomotive is running on the path~$I$ of the units
which is devoted to perform the incrementation. The path~$I$ leads the locomotive to
the \RR-gate of the \II-memory of the unit. Remember that both units contain the same
bit. 

If the locomotive reads~1{} in the \II-memory, it leaves the memory through its 
gate~\bbu{} so that the locomotive follows the continuation of the path~$I$ until
it arrives at the next unit. 

Accordingly, the locomotive crosses the units which contain~1 until it reaches \RRR($n$)
for the smallest~$n$ such that the unit contains~0. Consequently we are in the case 
when, entering the \II-memory through its \RR-gate, the locomotive reads~0, so that it 
leaves the memory through its gate~\bbz. There, a path leads the locomotive to the 
\WW-gate of the same memory, so that the locomotive rewrite its bit from~0 to~1. It exits
through the \EE-gate of the memory where it is led to a small disc~$s$ indicating a 
continuation through another~$s$ small disc of the right-hand side window. From that 
second~$s$, the locomotive goes to the \WW-gate of the \DD-memory, so that the second 
bit of the unit is turned from~0 to~1. Accordingly, when the locomotive leaves the 
\DD-memory through the \EE-gate, the unit is set to~1 which means that the 
incrementation is performed. From the \EE-gate, the locomotive is led to the 
$R$-path which is a return
path, leading the locomotive back to the \DDO-structure. From there, the locomotive
will reach the \DDI-structure which will send it back to the appropriate place in the
program.

\subsubsection{Decrementation}\label{sssdec}

Presently, consider the case of a locomotive coming to~\RRR{} in order to decrement it.
The locomotive arrives at the first unit through the path~$D$.

We first consider the case when the content~$c$ of~\RRR{} is positive. The locomotive 
arrives at~\RRR$(0)$ through a path~$D$ which we can see on the lower part of the
left-hand side part of Figure~\ref{funitreg}. The path is continued on the right-hand side
part and it leads the locomotive to the \RR-gate of the \DD-memory. As far as $c>0$,
the locomotive reads~1. Leaving the memory through its gate~\bbu, the locomotive is sent 
back to the $D$-path in order to go to visit the \DD-memory of the next unit.

\def\wdu{{\bf wd1}}
\def\wdt{{\bf wd2}}
Accordingly,the locomotive arrives at~\RRR($c$) whose content is~0 as well as the content
of~\RRR($n$) for \hbox{$n\geq c$}. It means that the locomotive has to go back to
\RRR($c$-1) in order to set the content of that unit to~0. To that goal, the locomotive
is sent from~\bbz{} to a path \wdu{} which leads the locomotive to the \WW-gate of the
\II-memory of \RRR($c$-1). So that the locomotive follows the same path as in the case
of an incrementation. Now, remember that entering a one-bit memory through its \WW-gate
rewrites its bit: if it is 0, it becomes~1, it is 1, it becomes~0. By assumption
the bits in the memories of \RRR($c$$-$1) are both set to~1 so that entering the 
\WW-gate of the \II-memory, the locomotive eventually leaves the \DD-memory through its 
\EE-gate which sends it to the return path $R$ and both memories of the unit are set to~0.
Consequently, in that case, the locomotive performed the decrementation. The 
\DDO-structure will send it to the \DDD-one through the $D$-path so that the locomotive 
will reach the program at the appropriate place.

Presently, consider the case when \RRR{} is empty. The locomotive still arrives at
the \RR-gate of the \DD-memory of the first unit of~\RRR. But reading~0, the locomotive
leaves the unit through the~\bbz-gate which sent it to an auxiliary path, the \wdu-one
which initiates the $Z$-path going to the \DDO-structure. The locomotive is sent through 
the $Z$-path to the \DDD-structure, so that it eventually reaches the appropriate place
of the program.

   Figure~\ref{funitreg} illustrates the arrival of the locomotive at the \WW-gate of
the \II-memory through another auxiliary path, \wdt, which comes from the unit \RRR($n$+1)
assuming that the figure illustrates \RRR$(n)$. Clearly, the incrementation is performed
on the unit where the locomotive reads~0, the decrementation is performed in the
preceding unit which explains what happens when the register is empty. Accordingly,
a successful decrementation implies that the locomotive goes back through the $R$-path
while an unsuccessful one implies the return through the \wdu-path: it is always 
possible to fix that one for the units with an even index while \wdt{} operates for
the units whose index is odd.

\section{The rules}\label{srules}

   In Sub-section~\ref{ssrlform}, we define the format of the rules and we discuss
the way we define the totalisticity. In Sub-section~\ref{ssrtracks} we give the rules
for the elements of the tracks, for the control and for the forks.

\subsection{Format of the rules and totalisticity}\label{ssrlform}

   A cell of the dodecagrid in the present cellular automaton consists of two elements:
a tile of the tiling, we call it the {\bf support} of the cell, and a finite automaton 
defined by the rules of the present section. Note that the finite automaton is the
same for all cells. The neighbours of a cell~$c$ whose support is~$\Delta$ are the cells 
whose support are the $\Delta_i$, \hbox{$i\in\{0..11\}$}, where $\Delta_i$ shares the 
face~$i$ of~$\Delta$. In that case, $\Delta_i$ is called the {\bf $i^{\rm th}$-neighbour} 
of~$c$, its {\bf $i$-neighbour} for short. The numbering of the faces is the one we 
described in Section~\ref{intro} with the help of Figure~\ref{fdodecs}.

   As usual, we call {\bf alphabet} of our cellular automaton the finite set of its 
possible states. In the present case, the alphabet of our cellular automaton consists 
of \sww, \sss, \srr{} and \sgg. We also call \sww{} the {\bf blank} as far as it is the 
state of the cells of the dodecagrid except finitely many of them. Let $c$ be a cell of 
the dodecagrid whose state is {\ftt S$_o$ } and whose support is $\Delta$, and let 
{\ftt S$_i$ } be the state of its $i$-neighbour.
Let {\ftt S$_n$ } be the {\bf new} state of~$c$, {\it i.e.} the state taken by~$c$
when it is the state associated with {\ftt S$_o$ } and the {\ftt S$_i$ } by the
automaton. We write this as follows:

\def\PP{$\mathbb P$}
\def\llrule #1 #2 #3 #4 #5 #6 #7 {%
\hbox{\ftt{#1.#2#3#4#5#6#7} }
}
\def\rrrule #1 #2 #3 #4 #5 #6 #7 {%
\hbox{\ftt{#1#2#3#4#5#6.#7} }
}
\vskip 5pt
\ligne{\hfill
\llrule {S$_o$} {S$_0$} {S$_1$} {S$_2$} {S$_3$} {S$_4$} {S$_5$} 
\hskip-4pt\rrrule {S$_6$} {S$_7$} {S$_8$} {S$_9$} {S$_{10}$} {S$_{11}$} {S$_n$} {}
\hfill$(1)$\hskip 10pt}
\vskip 5pt
\noindent
and we say that {\ftt S } is a {\bf rule} of the automaton. All rules of the automaton
we give in the present section obey the format defined by~$(1)$. Of course, {\ftt S$_o$ }
{\ftt S$_n$ } and the {\ftt S$_i$ } belong to the alphabet 
\hbox{$\{$\sww, \sss, \srr, \sgg$\}$}.

   Note that if $\sigma$ is a permutation on \hbox{$\{0..11\}$} we call {\bf permuted 
image of {\ftt S } under $\sigma$} the rule: 
\vskip 5pt
\ligne{\hfill 
\llrule {S$_o$} {S$_{\sigma(0)}$} {S$_{\sigma(1)}$} {S$_{\sigma(2)}$} {S$_{\sigma(3)}$} 
{S$_{\sigma(4)}$} {S$_{\sigma(5)}$} 
\hskip-4pt\rrrule {S$_{\sigma(6)}$} {S$_{\sigma(7)}$} {S$_{\sigma(8)}$} 
{S$_{\sigma(9)}$} {S$_{\sigma(10)}$} {S$_{\sigma(11)}$} {S$_n$} {} 
\hfill\hskip 10pt}
\vskip 5pt
\setbox120=\hbox{\ftt S }
\noindent
which we denote {\ftt  S }$_\sigma$. Let {\ftt S } be a rule, we call {\bf weight} of
{\ftt S } denoted by $w$({\ftt S }) the number 
$\displaystyle{\sum\limits_{i=0}^{11}w(\copy120_i)}$ where $w$({\ftt S }$_i$) is the
weight of the considered state {\it i.e.} the rank of that state in
\hbox{$\{$\sww,\sss,\srr,\sgg$\}$}, the rank of \sww{} being~0. Accordingly, in the line
giving a rule we also indicate its weight. As an example, the quiescent rule will be 
written as follows:

\def\hhzw{\hskip 5pt}
\def\hhwww{\hskip 20pt}
\def\hhww{\hskip 10pt}
\def\totregle o#1v#2n#3p#4 #5 {%
\hbox to 115pt {\hbox to 15pt{\hfill\ftt {#5 } }\hhzw\ftt {#1.#2.#3 } \hhzw {\ftt\bf #4 } 
\hfill}
}
\vskip 5pt
\ligne{\hfill\totregle o{W}v{WWWWWWWWWWWW}n{W}p{0} {1} \hfill}

It is the first rule and its weight is 0.

    As a last point, we shall present the rules not only by splitting their set according
to their role in the simulation but also, inside each of such subsets, according to the
role of the rule with respect to the motion itself. We noted in the previous sections
what we called idle configurations. In such a condition, the configuration must remain 
unchanged as long as the locomotive is not
in the window which defines the configuration. Such rules are called {\bf conservative}.
When the locomotive falls within the window, some cells are changed while the others
remain unchanged. Those which are changed are called {\bf motion rules}.

\subsection{The rules for the tracks}\label{ssrtracks}

   We start our study of the rules and the construction of the rules by those which
manage the tracks. As mentioned in Sub-subsection~\ref{ssstracks}, the motion of the
locomotive is depicted by the scheme~(2) we reproduce below for the convenience of the
reader.
\vskip 5pt
\ligne{\hfill\ftt{
   3B234234\hskip 20pt 34R34234\hskip 20pt 342G4234\hskip 20pt 3423B234\hskip 20pt 
3423B234\hskip 20pt 342342R4}
\hfill(2)\hskip 10pt}
\vskip 5pt
\def\fttd{{\ftt 2 }}
\def\fttt{{\ftt 3 }}
\def\fttq{{\ftt 4 }}
Here, \fttd, \fttt{} and \fttq{} denote elements of the tracks with 2, 3 and 4 milestones
respectively in their decoration. Figure~\ref{ftracks} shows us such elements.
The conservative rules induced by those elements are given by Table~\ref{trtrackscv}.
The first rules, thirteen of them, deal with the milestones. When the current state
is not \sww, it means that the locomotive is seen by the milestone. Note that \sss{} is 
not the state of a milestone.
\newcount\nbregle\nbregle=1

\vskip-10pt
\ligne{\hfill                                         
\vtop{\leftskip 0pt\parindent 0pt                     
\begin{tab}\label{trtrackscv}
\leurre
Conservative rules for the tracks. When a rule is repeated, for example rule~$6$, its
number is also repeated.
\end{tab}
\vskip-10pt
\ligne{\hfill milestones\hfill}                       
\ligne{\hfill                                         
\vtop{\leftskip 0pt\parindent 0pt\hsize=125pt         
\ligne{\totregle o{W}v{WWWWWWWWWWWW}n{W}p{0} {\the\nbregle} 
\global\advance\nbregle by 1\hfill}
\vskip-4pt
\ligne{\totregle o{G}v{WWWWWWWWWWWW}n{G}p{0} {\the\nbregle} 
\global\advance\nbregle by 1\hfill}
\vskip-4pt
\ligne{\totregle o{R}v{WWWWWWWWWWWW}n{R}p{0} {\the\nbregle} 
\global\advance\nbregle by 1\hfill}
\vskip-4pt
\ligne{\totregle o{W}v{RWWWWWWWWWWW}n{W}p{2} {\the\nbregle} 
\global\advance\nbregle by 1\hfill}
\vskip-4pt
\ligne{\totregle o{W}v{GWWWWWWWWWWW}n{W}p{3} {\the\nbregle} 
\global\advance\nbregle by 1\hfill}
\vskip-4pt
\ligne{\totregle o{W}v{GGWWWWWWWWWW}n{W}p{6} {\the\nbregle} 
\global\advance\nbregle by 1\hfill}
\vskip-4pt
\ligne{\totregle o{W}v{RGWWWWWWWWWW}n{W}p{5} {\the\nbregle} 
\global\advance\nbregle by 1\hfill}
}                                                     
\hskip 10pt
\vtop{\leftskip 0pt\parindent 0pt\hsize=125pt         
\ligne{\totregle o{R}v{BWWWWWWWWWWW}n{R}p{1} {\the\nbregle} 
\global\advance\nbregle by 1\hfill}
\vskip-4pt
\ligne{\totregle o{R}v{RWWWWWWWWWWW}n{R}p{2} {\the\nbregle} 
\global\advance\nbregle by 1\hfill}
\vskip-4pt
\ligne{\totregle o{R}v{GWWWWWWWWWWW}n{R}p{3} {\the\nbregle} 
\global\advance\nbregle by 1\hfill}
\vskip-4pt
\ligne{\totregle o{G}v{BWWWWWWWWWWW}n{G}p{1} {\the\nbregle} 
\global\advance\nbregle by 1\hfill}
\vskip-4pt
\ligne{\totregle o{G}v{RWWWWWWWWWWW}n{G}p{2} {\the\nbregle} 
\global\advance\nbregle by 1\hfill}
\vskip-4pt
\ligne{\totregle o{G}v{GWWWWWWWWWWW}n{G}p{3} {\the\nbregle} 
\global\advance\nbregle by 1\hfill}
}                                                     
\hfill}                                               
\vskip 4pt
\ligne{\hfill elements of the track\hfill}            
\ligne{\hfill                                         
\vtop{\leftskip 0pt\parindent 0pt\hsize=125pt         
\ligne{\totregle o{W}v{GGWWWWWWWWWW}n{W}p{6} {6} \hfill}
\vskip-4pt
\ligne{\totregle o{W}v{GGRWWWWWWWWW}n{W}p{8} {\the\nbregle} 
\global\advance\nbregle by 1\hfill}
\vskip-4pt
\ligne{\totregle o{W}v{GGGRWWWWWWWW}n{W}p{11} {\the\nbregle} 
\global\advance\nbregle by 1\hfill}
}                                                     
\hfill}                                               
\vskip 4pt
\ligne{\hfill control \hfill fork\hfill}              
\ligne{\hfill                                         
\vtop{\leftskip 0pt\parindent 0pt\hsize=125pt         
\ligne{\totregle o{W}v{GGGGGWWWWWWW}n{W}p{15} {\the\nbregle} 
\global\advance\nbregle by 1\hfill}
\vskip-4pt
\ligne{\totregle o{W}v{GGGGGGWWWWWW}n{W}p{18} {\the\nbregle} 
\global\advance\nbregle by 1\hfill}
\vskip-4pt
}                                                     
\hskip 10pt
\vtop{\leftskip 0pt\parindent 0pt\hsize=125pt         
\ligne{\totregle o{W}v{GGWWWWWWWWWW}n{W}p{6} {6} \hfill}   
}                                                     
\hfill}                                               
}                                                     
\hfill}                                               
\vskip 15pt

Note that rule~6 is repeated: it first occurs for a blank tile which can see two
\sgg-milestones when they are put on adjacent faces of an element of the tracks.
In the second occurrence of the rule, it is an element of the track, namely the
element {\ftt 2 } as far as its decoration consists of two milestones. In the table,
rules~2 up to~5 say that a blank neighbour of a milestone remains blank while 
a milestone itself whose all neighbours are blank does not change its state.
Rules~6 and~7 deal with a blank tile which can see two milestones: either two \sgg-ones
or a \sgg-one together with an \srr-one. Rules~8 up to~10 say that an \srr-milestone is
not affected when it can see a locomotive, whichever its state. The same role is
performed by rules~11 up to~13 for a \sgg-milestone.

\vskip-10pt
\ligne{\hfill                                         
\vtop{\leftskip 0pt\parindent 0pt                     
\begin{tab}\label{trtracksmv}
\leurre
Motion rules for the tracks. 
\end{tab}
\vskip-10pt
\ligne{\hfill the tracks\hfill}                       
\ligne{\hfill                                         
\vtop{\leftskip 0pt\parindent 0pt\hsize=125pt         
\ligne{\hfill {\ftt 2 } \hfill}
\ligne{\totregle o{W}v{GGWWWWWWWWWW}n{W}p{6} {6} \hfill} 
\vskip-4pt
\ligne{\totregle o{W}v{GGWBWWWWWWWW}n{R}p{7} {\the\nbregle} 
\global\advance\nbregle by 1\hfill}
\vskip-4pt
\ligne{\totregle o{R}v{GGWWWWWWWWWW}n{W}p{6} {\the\nbregle} 
\global\advance\nbregle by 1\hfill}
\vskip-4pt
\ligne{\totregle o{W}v{GGWGWWWWWWWW}n{W}p{9} {\the\nbregle} 
\global\advance\nbregle by 1\hfill}
}                                                     
\hskip 10pt
\vtop{\leftskip 0pt\parindent 0pt\hsize=125pt         
\ligne{\hfill {\ftt 3 } \hfill}
\ligne{\totregle o{W}v{GGRWWWWWWWWW}n{W}p{8} {14} \hfill} 
\vskip-4pt
\ligne{\totregle o{W}v{GGRWRWWWWWWW}n{G}p{10} {\the\nbregle} 
\global\advance\nbregle by 1\hfill}
\vskip-4pt
\ligne{\totregle o{G}v{GGRWWWWWWWWW}n{W}p{8} {\the\nbregle} 
\global\advance\nbregle by 1\hfill}
\vskip-4pt
\ligne{\totregle o{W}v{GGRWBWWWWWWW}n{W}p{9} {\the\nbregle} 
\global\advance\nbregle by 1\hfill}
}                                                     
\hfill}                                               
\vskip 4pt
\ligne{\hfill                                         
\vtop{\leftskip 0pt\parindent 0pt\hsize=125pt         
\ligne{\hfill {\ftt 4 } \hfill}
\ligne{\totregle o{W}v{GGGRWWWWWWWW}n{W}p{11} {15} \hfill} 
\vskip-4pt
\ligne{\totregle o{W}v{GGGRWGWWWWWW}n{B}p{14} {\the\nbregle} 
\global\advance\nbregle by 1\hfill}
\vskip-4pt
\ligne{\totregle o{B}v{GGGRWWWWWWWW}n{B}p{11} {\the\nbregle} 
\global\advance\nbregle by 1\hfill}
\vskip-4pt
\ligne{\totregle o{W}v{GGGRWRWWWWWW}n{W}p{13} {\the\nbregle} 
\global\advance\nbregle by 1\hfill}
}                                                     
\hskip 10pt
\vtop{\leftskip 0pt\parindent 0pt\hsize=125pt         
\ligne{\hfill {\ftt fork } \hfill}
\ligne{\totregle o{W}v{GGWWWWWWWWWW}n{W}p{6} {6} \hfill} 
\vskip-4pt
\ligne{\totregle o{W}v{GGWBWWWWWWWW}n{R}p{7} {19} \hfill} 
\vskip-4pt
\ligne{\totregle o{R}v{GGWWWWWWWWWW}n{W}p{6} {20} \hfill}
\vskip-4pt
\ligne{\totregle o{W}v{GGWGGWWWWWWW}n{W}p{12} {\the\nbregle} 
\global\advance\nbregle by 1\hfill}
}                                                     
\hfill}                                               
\vskip 4pt
\ligne{\hfill control\hskip 90pt its access\hfill}    
\ligne{\hfill                                         
\vtop{\leftskip 0pt\parindent 0pt\hsize=125pt         
\ligne{\totregle o{W}v{GGGGGWWWWWWW}n{W}p{15} {16} \hfill} 
\vskip-4pt
\ligne{\totregle o{R}v{GGGGGWWWWWWW}n{R}p{15} {\the\nbregle} 
\global\advance\nbregle by 1\hfill}
\vskip-4pt
\ligne{\totregle o{W}v{GGGGGWRWWWWW}n{R}p{17} {\the\nbregle} 
\global\advance\nbregle by 1\hfill}
\vskip-4pt
\ligne{\totregle o{R}v{GGGGGWRWWWWW}n{W}p{17} {\the\nbregle} 
\global\advance\nbregle by 1\hfill}
\ligne{\totregle o{W}v{GGGRWRGWWWWW}n{W}p{16} {\the\nbregle} 
\global\advance\nbregle by 1\hfill}
}                                                     
\hskip 10pt
\vtop{\leftskip 0pt\parindent 0pt\hsize=125pt         
\ligne{\totregle o{W}v{GGGGGGWWWWWW}n{W}p{18} {17} \hfill} 
\vskip-4pt
\ligne{\totregle o{W}v{GGGGGGWBWWWW}n{R}p{19} {\the\nbregle} 
\global\advance\nbregle by 1\hfill}
\vskip-4pt
\ligne{\totregle o{R}v{GGGGGGWWWWWW}n{W}p{18} {\the\nbregle} 
\global\advance\nbregle by 1\hfill}
\vskip-4pt
\ligne{\totregle o{R}v{GGGGGGWRWWWW}n{W}p{20} {\the\nbregle} 
\global\advance\nbregle by 1\hfill}
\ligne{\totregle o{G}v{WGRWWWWWWWWW}n{G}p{5} {\the\nbregle} 
\global\advance\nbregle by 1\hfill}
}                                                     
\hfill}                                               
}                                                     
\hfill}                                               
\vskip 10pt
Table~\ref{trtracksmv} gives all the motion rules involved in the computation. In order
to better understand the rules, the conservative rules of each configuration has been
repeated: it can be recognised by the fact that its number is the same as the one it
received in Table~\ref{trtrackscv}. Moreover, in the neighbourhoods of the rules,
we first give the decorations and then, separated from them by a \sww-symbol, the
possible occurrences of non-blank states. They can be the state of a locomotive, in
the considered element of the track. They can be two non-blank states in the case of
the controller or in the case of the fork.

The reader may notice that the rules for the motion in an element of the track are
parallel in the different cases: rules~6, 14 and~15 are the conservative rules in an
idle configuration. The way the locomotive crosses an element is given by~(4):
\vskip 5pt
\ligne{\hfill
$\vcenter{\vtop{\leftskip 0pt\parindent 0pt\hsize=210pt
\ligne{\hbox to 35 pt{\ftt element \hfill}
\hbox to 50pt{\hfill\ftt rules \hfill}\hbox to 35pt{\hfill\ftt entry \hfill}
\hbox to 35pt{\ftt inside \hfill}\hbox to 35pt{\hfill\ftt exit \hfill} \hfill}
\ligne{\hbox to 35 pt{\hfill\ftt 2 \hfill}
\hbox to 50pt{\hfill\ftt {18, 19, 20 } \hfill}\hbox to 35pt{\hfill\sss \hfill}
\hbox to 35pt{\hfill\srr \hfill}\hbox to 35pt{\hfill\sgg \hfill} \hfill}
\ligne{\hbox to 35 pt{\hfill\ftt 3 \hfill}
\hbox to 50pt{\hfill\ftt {21, 22, 23 } \hfill}\hbox to 35pt{\hfill\srr \hfill}
\hbox to 35pt{\hfill\sgg \hfill}\hbox to 35pt{\hfill\sss \hfill} \hfill}
\ligne{\hbox to 35 pt{\hfill\ftt 4 \hfill}
\hbox to 50pt{\hfill\ftt {24, 25, 26 } \hfill}\hbox to 35pt{\hfill\sgg \hfill}
\hbox to 35pt{\hfill\sss \hfill}\hbox to 35pt{\hfill\srr \hfill} \hfill}
}}$
\hfill(\numerrel)\hskip 10pt}
\vskip 5pt
Note that the fork is a {\ftt 2 }-element for which rules 6, 18 and 19 apply. Now,
instead of rule~20, rule 27 applies as far as two locomotives leave the element.
It is witnessed by rule~27 which indicates two \sgg-states outside the states 
corresponding to its decoration.

The access to the controller is dealt with by rule 17, the conservative rule, together
with rules~32 up to~34. A \sss-locomotive makes the cell becomes \srr, which allows
the controller to change its state: from \sww{} to \srr{} by rule~29 and from \srr{}
to \sww{} by rule~30. Note that rule~16 is the conservative rule for a blank controller
while rule~28 conservatively applies when it is \srr. Rule~31 shows us that the locomotive
when the element is {\ftt 4 }, a \sgg-locomotive is stopped by an \srr-controller.
At last, rule~35 is a conservative rule for a \sgg-milestone of a controller or of
its access cell when an \srr-locomotive crosses the considered cell. Indeed, as can be
seen on Figure~\ref{fctrl}, each one of two milestones of the controller can see
a milestone of the access to the controller and, of course, conversely. So that rule~35
applies both to the controller and to its access cell.

Note that we fixed the controller to be placed under a {\ftt 4 }-element, as shown by
rule~31. The rule says that the locomotive, which is then in the \sgg-state, cannot enter 
the element. That requires the arrival to the access cell of the controller to be the
case of a \sss-locomotive. If it is not provided by the standard periodic motion along
a track, that can be arranged through a path as indicated in the passive fixed switch or
in the fork.

\vskip 10pt
   That last point completes the proof of Theorem~\ref{letheo}. \hfill$\Box$

   Figure~\ref{fmvtracks} illustrates the motion of the locomotive on a portion of
a path. Note that the path is rather a simple one: it avoids a single cell of a track 
which requires three additional cells. In the figure, the first element of the appended
cells is a {\ftt 2 }-one so that the \sss-locomotive leaving the track became an 
\srr-locomotive when it is again on the track. It is conformal to the periodic motion
on a track. Note that the configuration of the milestones of the {\ftt 3 }-element of 
the path is different from those of the track: indeed, in another disposition, we would
have milestones of a {\ftt 3 }-element which would see a milestone of a neighbouring
{\ftt 4 }-element. It is avoided so that the rules apply without any problem. For a 
similar reason, in the decoration of the {\ftt 2 }-element of the track at which the 
path arrives, we place the milestones so that none of them can see a milestone of a
neighbouring element. We place one milestone on face~11 and the other on a face of the
upper crown which points at a blank neighbour of the cell in \HH$_u$.

\vskip 10pt
\vtop{
\ligne{\hfill
\includegraphics[scale=0.65]{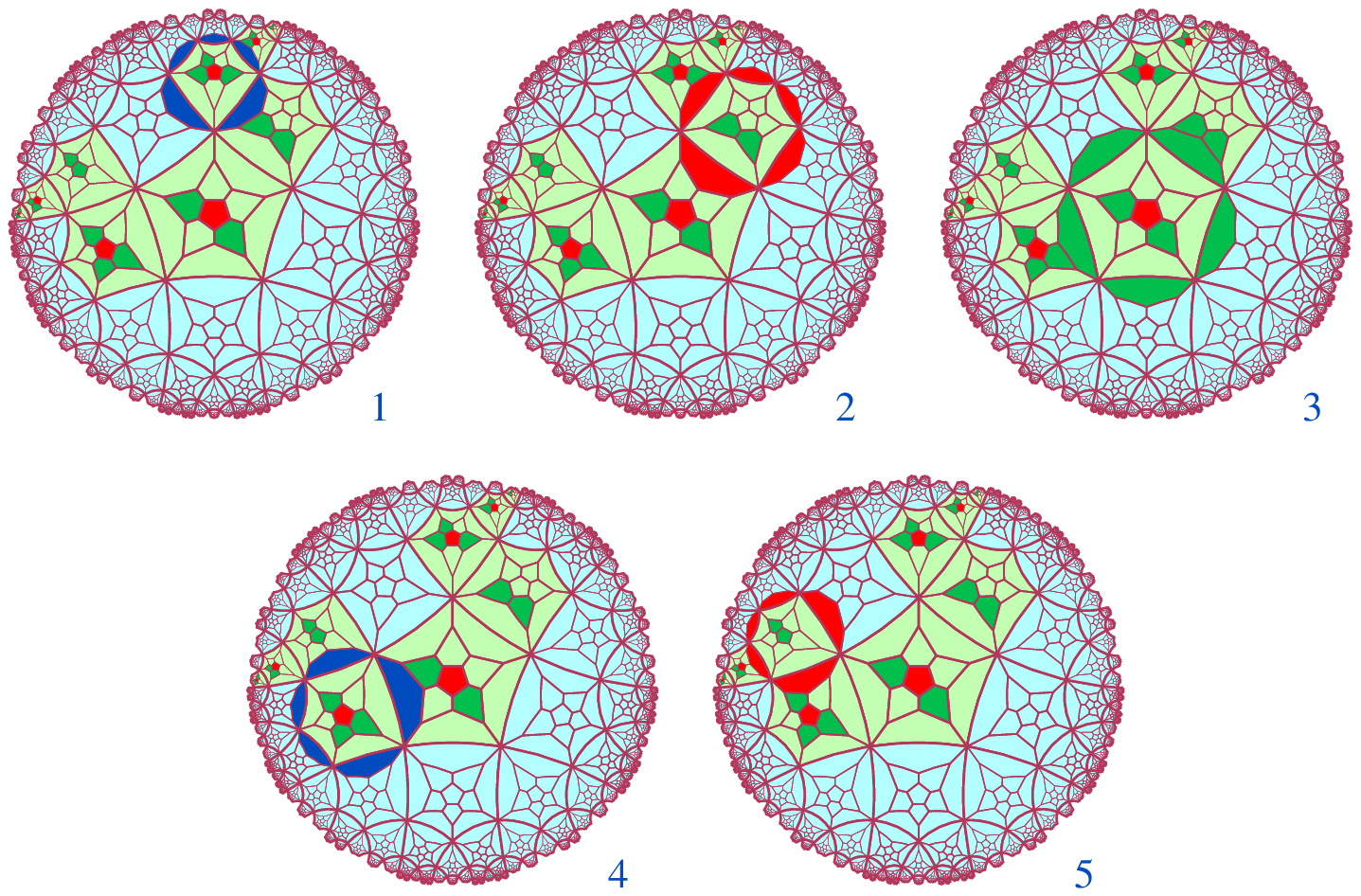}
\hfill}
\begin{fig}\label{fmvtracks}
\leurre
The motion of the locomotive on a portion of a path constructed for the purpose.
Note the successive states taken by the locomotive. When the locomotive is in a cell~$c$
we can see its colour on the faces of the neighbours of~$c$ which are adjacent to
the face~$0$ of~$c$. The indices show us the order of the pictures.
\end{fig}
}
\section{Conclusion}

    We have 35 rules for our cellular automaton which is outer totalistic and weakly 
universal. There is no rule which is used with the same weight and the same current 
state in different neighbourhoods. Several rules occur with the same weight and with 
\sww{} as the current state: they have different contexts. It occurs with the weights 9 
and~12. The present cellular automaton is not completely totalistic: rules~5, 9 and~13 
have a total weight of~3, taking the weight of the current state into account and their 
new state is different. Another example is given by rules~25 and~27.

   Four states seem to be needed for the motion of the locomotive on the tracks, as
already noticed. However, other results with cellular automata on which a different 
constraint is put might be obtained. It is an open question. Another open question is 
to consider completely totalistic cellular automata. Another one is to look whether it 
is possible to construct an outer totalistic cellular automaton which would also be 
strongly universal, {\it i.e.} whose set of non-blank cells would always be finite.

And so, there are a lot of open questions.

\end{document}